\documentclass[aps,pra,reprint,superscriptaddress,nofootinbib]{revtex4-1}

\usepackage{graphicx}
\usepackage{xcolor}
\usepackage{amsmath,esint}
\usepackage{amssymb,bm}
\usepackage{braket}
\usepackage{enumitem}
\usepackage{float}
\usepackage{tikz}
\usepackage[colorlinks=true,linkcolor=blue,citecolor=red]{hyperref}
\usepackage{mathtools}
\usepackage{tcolorbox}
\usepackage{relsize}
\usepackage{multirow}

\newcommand{\Li}{\mathcal{L}}

\newcommand{\rmd}{\text{d}}

\newcommand{\nS}{n_{\textsc{s}}}
\newcommand{\nt}{n_{\textsc{t}}}

\newcommand{\ws}{w_{\textsc{s}}}
\newcommand{\wt}{w_{\textsc{t}}}
\newcommand{\Gs}{\Gamma_{\textsc{s}}}
\newcommand{\Gt}{\Gamma_{\textsc{t}}}
\newcommand{\eref}[1]{(\ref{#1})}
\newcommand{\ko}{\mathcal{K}}
\newcommand{\fo}{\mathcal{F}_{1}}
\newcommand{\ft}{\mathcal{F}_{2}}

\newcommand{\M}{m'}
\graphicspath{{./figures/}}

\newcommand{\be}{\begin{equation}}
\newcommand{\ee}{\end{equation}}
\newcommand{\ben}{\begin{eqnarray}}
\newcommand{\een}{\end{eqnarray}}
\newcommand{\bF}{\begin{figure}}
\newcommand{\eF}{\end{figure}}
\newcommand{\bi}{\begin{itemize}}
\newcommand{\ei}{\end{itemize}}
\newcommand{\OoM}[1]{\mathcal{O}\left( #1 \right)}

\makeatletter
\newcommand{\subalign}[1]{%
  \vcenter{%
    \Let@ \restore@math@cr \default@tag
    \baselineskip\fontdimen10 \scriptfont\tw@
    \advance\baselineskip\fontdimen12 \scriptfont\tw@
    \lineskip\thr@@\fontdimen8 \scriptfont\thr@@
    \lineskiplimit\lineskip
    \ialign{\hfil$\m@th\scriptstyle##$&$\m@th\scriptstyle{}##$\hfil\crcr
      #1\crcr
    }%
  }%
}
\makeatother

\begin{document}

\title{Extracting electromagnetic signatures of spacetime fluctuations}

\author{B.~Sharmila}
\email{Sharmila.Balamurugan@warwick.ac.uk}
\affiliation{Department of Physics, University of Warwick, Coventry CV4 7AL, UK.}

\author{Sander M. Vermeulen}
\email{smv@caltech.edu}
\affiliation{School of Physics and Astronomy, Cardiff University, UK.}
\affiliation{Division of Physics, Mathematics and Astronomy,
California Institute of Technology, Pasadena, CA 91125, USA.}

\author{Animesh Datta}
\email{Animesh.Datta@warwick.ac.uk}
\affiliation{Department of Physics, University of Warwick, Coventry CV4 7AL, UK.}

\date{\today}

\begin{abstract}

We present a formalism to discern the effects of fluctuations of the spacetime metric on electromagnetic radiation. 
The formalism works via the measurement of electromagnetic field correlations, while allowing a clear assessment of the assumptions involved. As an application of the formalism, we present a model of spacetime fluctuations that appear as random fluctuations of the refractive index of the vacuum in single, 
and two co-located Michelson interferometers. We compare an interferometric signal predicted using this model to experimental data from the Holometer and aLIGO. We show that if the signal manifests at a frequency at which the interferometers are sensitive, the strength and scale of possible spacetime fluctuations can be constrained. The bounds, thus obtained, on the strength and scale of the spacetime fluctuations, are also shown to be more stringent than the bounds obtained previously using astronomical observation at optical frequencies. 
The formalism enables us to evaluate proposed experiments such as QUEST for constraining quantum spacetime fluctuations and to design new ones.

\end{abstract}

\keywords{spacetime fluctuations, electromagnetic field correlations, interferometers}

\maketitle

\section{\label{sec:intro}Introduction}

The theoretical outcome of combining quantum mechanics and general relativity remains one of the open problems of physics.
In 1957, Wheeler hypothesised a fluctuating spacetime metric~\cite{wheeler1957} while Wigner and Salecker noted the 
existence of fundamental limits on the measurement of spacetime distances~\cite{Wigner1957,Salecker1958,Hossenfelder2013}.
Conceptual aspects of such fluctuations and the measurability of the gravitational field were also discussed at the 1957 Chapel Hill Conference~\cite{Chapel1957}.
However, it was not until 1999 that an experimental route to observe such fluctuations using laser-interferometric gravitational wave detectors was suggested~\cite{amelino99}.
Complementary to terrestrial detectors, astronomical observations were also suggested for similar purposes~\cite{AmelinoCamelia1998}. 

The general strategy has since been to constrain hypothesised theoretical models using data from these detectors and observations~\cite[Table 1]{NgvanDam2000}.
The former motivated the derivation of a device-independent limit on the minimum length that can be measured~\cite{Calmet2004}.
The latter includes a variety of sources whose emissions have been exploited to constrain proposed models~\cite{Perlman2015,NgPerlman2022}. 
The increasing sensitivity of laser-interferometric gravitational wave detectors has also placed constraints on some of these models.
Indeed, the data from dedicated co-located and co-aligned Michelson interferometers have also been used to rule out certain hypothesised non-commutative and holographic models of spacetime~\cite{Chou2017,bentHolo}.

Models based on the hypothesised holographic principle of quantum gravity have, in particular, received some recent attention. 
While certain ways of invoking this holographic principle have seemingly been ruled out by astronomical observations~\cite{NgPerlman2022},
others have been studied for their signatures in Michelson interferometers~\cite{VerZurek21,Zurek2022,geontropy2023}.   
These single-parameter models seek to provide a microscopic mechanism for spacetime fluctuations, and 
their validity is to be probed in forthcoming experiments with Michelson interferometers such as QUEST and GQuEST~\cite{cardiff21,QUESTconf,mcculler}.

In this paper, we present a formalism that describes how fluctuations in the spacetime metric manifest as measurable signatures in electromagnetic radiation propagating through it.
The fluctuations in the elements of the metric tensor are modelled using their strengths and scales. 
Signatures of such fluctuations are identified in general electromagnetic field correlation tensors after propagation.
We illustrate the strength of our formalism by constraining models of spacetime fluctuations, in two-dimensional exclusion plots,
using the experimental data from a single as well as two co-located Michelson interferometers 
such as the Holometer~\cite{int1,HoloData} and aLIGO~\cite{LIGOdata}. The experimental data from these interferometric setups typically capture the fluctuations in the output beam power from its average value. In our approach, we obtain these output power fluctuations using the above-mentioned electromagnetic field correlation tensors, and a subsequent comparison with the experimental data helps in constraining the spacetime fluctuations.
We also apply our methods to the proposed QUEST~\cite{QUESTconf} experiment at the Cardiff University~\cite{cardiff21},
and relate our constraints to those obtained using some previous single-parameter models. 

In principle, a generic random process could require infinitely many parameters to completely describe its infinitely many moments. However, our two-parameter model is motivated by the empirical observation that physical phenomena may typically be characterised solely by their strength and scale. These are the two parameters captured in our model. A single-parameter model~\cite{Perlman2015,NgPerlman2022,geontropy2023} elides these two distinct physical aspects. In fact, the single-parameter models assume the spacetime fluctuations to be white noise, thereby implicitly assuming the scale to be zero~\cite{VerZurek21} or Planck length~\cite{Perlman2015,NgPerlman2022}. Evidently, the single-parameter models lose a degree of freedom by assuming the scale of the fluctuation to be a fixed value. Foregoing these implicit assumptions with our two-parameter model can thus become less restrictive and lead to a more faithful reflection of the physical universe. 
Using both the strength and the scale to model the metric fluctuations, in contrast to prior single-parameter models, 
enables stricter constraints from the cross-correlation between the outputs of the two co-located Michelson interferometers. 

We list the mathematical assumptions invoked for analytical tractability, and discuss their consequences. 
Indeed, invoking a subset of these assumptions even allows these fluctuations to be interpreted as spatial and temporal fluctuations in the refractive index of the medium.

Our work complements prior ones~\cite{kwon2022,geontropy2023,NgPerlman2022} by taking a phenomenological approach to relate spacetime fluctuations to their electromagnetic signatures.
Our model is not derived from a fundamental quantum theory of gravity and is agnostic of the underlying mechanism of spacetime fluctuations.
It rather captures the scenario where any residual fluctuations in spacetime measurements, after identifying known sources,
may be ascribed to possible quantum origins. 
Operationally, it allows the strength and scale of spacetime fluctuations to be constrained by experimental data.
This in turn helps our formalism provide an unambiguous way of judging the performance of experiments designed to constrain spacetime fluctuations and in formulating new experiments.
Finally, it should also enable a unified treatment for constraining spacetime fluctuations from terrestrial experiments and astronomical observations.

More generally, our work should enable unification with explorations of other consequences on spacetime fluctuations, particularly on massive~\cite{Karolyhazy66}  and charged objects~\cite{Fu1997}. 
The former is hypothesised to lead to spatial decoherence, and has already been constrained using interferometric gravitational wave detectors~\cite{collapseBnds} and parts thereof~\cite{collapseLISA}. 
The latter suggests the emission of radiation due to fluctuating accelerations, and has also been constrained using terrestrial experiments and astronomical observations~\cite{figurato2023}. 
These efforts are part of a larger body of work studying the coupling of quantum matter with gravity~\cite{Donadi2022}. 

The paper is structured as follows. 
Section \ref{sec:model} lists the attributes of our model of spacetime fluctuations. 
Section \ref{sec:EMsig} uses these attributes to simplify the relativistic wave equation for the propagation of 
light in curved spacetime. Their solution, subject to certain assumptions, provides analytical expressions for the relevant correlation tensors of the electric field. 
Section \ref{sec:IntSig} specialises the correlation functions to interferometric setups to obtain the spectral densities of the relevant observable in a single, and two co-located interferometers. 
These expressions are compared with the experimental data in Sec. \ref{sec:SDs} to obtain two-dimensional exclusion plots constraining the scale and strength of the fluctuations.
We use the experimental data from the Holometer experiment and aLIGO, as well as the proposed QUEST experiment. 
Section~\ref{sec:alphaComp} relates our results to prior constraints obtained using single-parameter models of spacetime fluctuations.
 We conclude in Sec.~\ref{sec:disc}.

\section{\label{sec:model} A Model of spacetime fluctuations}

We propose a phenomenological model to capture fluctuations in the spacetime metric tensor $g^{\mu \nu}.$
We begin by  listing the attributes of these fluctuations. 
Of the ten independent coefficients $g^{\mu\nu}$ that can fluctuate subject to symmetry, we focus on one ($g^{00}$) to illustrate our formalism. 
Some of these attributes are invoked to capture empirical insights while others to enable a tractable analysis. 
Each of these can be modified or revoked to further the understanding of possible spacetime fluctuations.

We consider metric fluctuations that are
\begin{enumerate}[label=(\arabic*)]
\item \label{mod:metric} of the form,
\begin{eqnarray}
g^{00}=- n^{2}(\bm{r}), \quad g^{ij}=\delta_{ij}, \quad g^{0i}=g^{i0}=0,
\end{eqnarray}
for $i,j=1,2,3$
where\footnote{We shall often label the components of 3-vectors by $x$, $y$, and $z$ instead of the tensor labels $1$, $2$, and $3$ respectively. } $n(\bm{r}),$ $\bm{r} \equiv (x,y,z,t)$ is a real function of its arguments\footnote{Metric fluctuations with $g^{00}=-1, \: g^{ij}=\frac{1}{n^{2}(x,y,z,t)}\delta_{ij}$, or $g^{00}=-n(x,y,z,t), \: g^{ij}=\frac{1}{n(x,y,z,t)}\delta_{ij}$ are equivalent.}.
The quadratic form of $g^{00}$ is not essential, but enables an appealing interpretation in Sec.~\ref{sec:EMsig}.

\item \label{mod:SmFl} such that $n(x,y,z,t)=\nS(x,y,z) \nt(t),$
where $\nS(x,y,z),\nt(t)$ are real functions of their arguments.
This attribute is crucial for variable separation assumed in Sec.~\ref{sec:EMsig}. 

\item \label{mod:small} small, that is, $\nS(x,y,z)=1 + \ws(x,y,z)$, $\nt(t)=1 + \wt(t)$, and $\ws, \wt \ll 1$ 
are real functions of their arguments. This captures the empirical fact that any fluctuations of spacetime are yet to be detected in the most sensitive of experiments.
It is invoked towards the end of Sec.~\ref{sec:EMsig}.

\item \label{mod:stat} zero mean, stationary random processes, that is, 
\begin{equation}
\overline{w_{\textsc{x}}}=0 \quad (\textsc{x=s,t}),
\end{equation}
\begin{equation}
\overline{\ws(x,y,z) \ws(x+\delta_{x},y+\delta_{y},z+\delta_{z})}\\
 = \Gs \, \rho(\delta_{x},\delta_{y},\delta_{z}),
 \label{eqn:spFluctCorr}
\end{equation}
and
\begin{equation}
\overline{\wt(t) \wt(t+\delta_{t})} = \Gt \, \varrho(\delta_{t}),
\end{equation}
where $\overline{\mathcal{O}}$ denotes the average of any quantity $\mathcal{O}$ over realisations of the random process.

\item \label{mod:indep} uncorrelated in space and time, that is, the average 
\begin{equation}
\overline{\ws(x,y,z)\,\wt(t)}= \overline{\ws(x,y,z)} \: \: \overline{\wt(t)}.
\end{equation}
Attributes \ref{mod:stat} and \ref{mod:indep} enable analytic computation of the relevant observables in Sec.~\ref{sec:SDs}.

\item \label{mod:gauss} Gaussian in space, that is,
\begin{equation}
\rho(\delta_{x},\delta_{y},\delta_{z})= \exp \left( -\sum\limits_{i=x,y,z} \dfrac{\delta_{i}^{2}}{2 \ell_{i}^{2}} \right),
\end{equation}
with correlation lengths $\ell_{i}, i=x,y,z.$ 
This attribute allows analytical closed-form expressions of relevant observables such as the power spectral and cross-spectral densities 
in Sec. \ref{sec:SDs}. 
We leave the form of $\varrho(\delta_{t})$ unspecified, noting that it may have a characteristic time scale $\tau_t$.

\end{enumerate}

Our model is thus parameterised only by the strength $\Gs$ and scale $ N_i  \equiv  2\ell_i/\ell_{\textsc{p}}$
of the fluctuations, 
where the Planck length $\ell_{\textsc{p}}$ 
is introduced to enable a fully dimensionless parameterisation.
The factor of 2 in the parametrisation of $N_i$ is chosen in anticipation of Sec. \ref{sec:SDs} 
where light propagates along the two arms of a Michelson interferometer oriented along the $x$ and $z$ axes, 
and the relevant observables become proportional to $(\ell_x + \ell_z)$.  
The observables will then be proportional to the chosen scale, 
especially in an isotropic model where $N_{\parallel} \equiv N_{x} = N_{y}= N_{z}$.

In what follows, we first describe how the spacetime fluctuations affect the propagation of electromagnetic fields. Using data from optical experiments, we then set upper bounds on the strength and scale of the spacetime fluctuations.

\section{\label{sec:EMsig} Electromagnetic signatures}

The relativistic wave equation governing the propagation of light in curved spacetime is~\cite{Friedlander1976,Tsagas2005} 
\begin{equation}
\Box F_{\alpha \beta} + 2 R_{\alpha \beta \gamma \delta} F^{\gamma \delta} - R_{\alpha \gamma} F^{\gamma}_{\beta} + R_{\beta \gamma} F^{\gamma}_{\alpha} =0,
\label{eqn:relativWE}
\end{equation}
where $F_{\alpha \beta}$ is the electromagnetic field tensor, $\Box$ denotes the d'Alembertian operator with the covariant derivative, $R_{\alpha \beta \gamma \delta}$ is the Riemann tensor, and $R_{\alpha \beta}$ is the Ricci curvature tensor defined in terms of the spacetime metric $g^{\mu \nu}$. 
Equation~\eref{eqn:relativWE} is obtained from Maxwell's equations written in terms of the covariant derivative without further assumptions of gauge or form of the 4-potential~\cite{Tsagas2005}. 

In principle, all observable signatures  of the propagation of electromagnetic fields through the fluctuating spacetime, can 
be extracted by solving Eq.~\eqref{eqn:relativWE} for a general tensor $g^{\mu\nu}.$
In practice, this seems intractable, even when the metric fluctuations are restricted to possess the six Attributes listed in Sec.~\ref{sec:model}.
Further assumptions must thus be made to solve Eq.~\eqref{eqn:relativWE}.

The Christoffel symbols contained in Eq.~\eqref{eqn:relativWE} involve derivatives of the metric tensor $g^{\mu\nu}.$ 
In particular, it contains derivatives of $g^{00}=- n^{2}(\bm{r})$, as per attribute \ref{mod:metric} of our model in Sec.~\ref{sec:model}.
The corresponding wave equation, provided in Appendix~\ref{app:one}, eluded our attempts at a closed-form analytical solution. 
To advance we thus assume that for a field tensor $F_{\alpha \beta}$, 
\begin{subequations}
\begin{eqnarray}
 \OoM{ \partial_{\mu} n^{2} } &\ll& \frac{ \OoM{\partial_{\nu} F_{\alpha \beta}} }{\OoM{ F_{\varphi \gamma} }}, \\
\quad \OoM{ \partial_{\mu} n^{2} } &\ll& \frac{\OoM{ \partial_{\chi} \partial_{\nu} F_{\alpha \beta} }}{\OoM{ \partial_{\sigma} F_{\varphi \gamma} }}, \quad\\
 \OoM{ \partial_{\mu} \partial_{\nu} n^{2} } &\ll& \frac{\OoM{ \partial_{\chi} \partial_{\sigma} F_{\alpha \beta} }}{\OoM{F_{\varphi \gamma} }},
\end{eqnarray}
\label{eqn:fluctIneq}
\end{subequations}
where $\OoM{\cdot}$ denotes an order-of-magnitude of the tensor quantity within the parentheses. 
These conditions suppose that any variation in $n^{2}$ across spacetime occurs at a rate much slower than the variation 
of the electromagnetic field.
Therefore, in subsequent analysis, the wavelength and time period of the wave, that set the natural scales of the electromagnetic field variations, should be much smaller than the corresponding scales for variations\footnote{In Sec. \ref{sec:SDs}, when assuming random fluctuations with the scale $N_{\parallel}$, this condition translates to $N_{\parallel} \gg \lambda/\ell_{\textsc{p}}$, where $\lambda$ is the wavelength of the light.} in $n^2$. 

The inequalities in \eref{eqn:fluctIneq} imply that the terms in Eq.~\eqref{eqn:relativWE} comprising either the Christoffel symbols or any derivative thereof
are negligible compared to $g^{\gamma \delta} \partial_{\delta} \partial_{\gamma} F_{\alpha \beta}$. 
This simplifies Eq.~\eqref{eqn:relativWE} to 
\begin{eqnarray}
g^{\gamma \delta} \partial_{\delta} \partial_{\gamma} F_{\alpha \beta}=0,    
\end{eqnarray}
which reduces to
\begin{subequations}
\begin{eqnarray}
\nonumber \nabla^{2} \mathbf{E} &=& \frac{n^{2}(x,y,z,t)}{c^{2}} \, \frac{\partial^{2} \mathbf{E}}{\partial t^{2}} \\ 
&& \text{for } \alpha=0, \, \beta=i \text{ or } \alpha=i, \, \beta=0, 
\label{eqn:SWE}\\
\nonumber \nabla^{2} \mathbf{B} &=& \frac{n^{2}(x,y,z,t)}{c^{2}} \, \frac{\partial^{2} \mathbf{B}}{\partial t^{2}}\\ 
&&\text{for } \alpha=i,\, \beta = j \neq i \,(i,j=1,2,3), 
\end{eqnarray}
\label{eqn:SWEpair}
\end{subequations}
where $\mathbf{E}$ and $\mathbf{B}$ denote 3-vector electric and magnetic fields. 
Using the Attribute \ref{mod:metric} and the inequalities~\eqref{eqn:fluctIneq}, the fluctuating metric tensor thus appears as an effective refractive index for the electromagnetic wave equation in free, flat space.
It is also evident from Eq. \eref{eqn:SWEpair} that the choice of the metric in Attribute \ref{mod:metric} is crucial for interpreting $n(\bm{r})$ as a 
refractive index fluctuating randomly in space and time. Propagation of light through a medium with homogeneous isotropic fluctuations of the refractive index was studied in Ref.~\cite{RandMedia68}. The stochastic nature of the \textit{random medium} is shown to manifest itself in the average of the fourth-order correlation of the radiation field over an ensemble of possible fluctuations of the refractive index.

Motivated by Ref.~\cite{RandMedia68}, we examine general electric field correlation tensors for observable signatures of the underlying spacetime fluctuations. Electric field correlation tensors can be of the form~\cite{MandelRMP,MandelWolf} 
\begin{widetext}
\be
M^{m,\M}_{\bm{c};\bm{c'}} \left(\{\bm{r}\}; \{\bm{r'}\} \right)
=
E_{c_1}(\bm{r}_1)
E_{c_2}(\bm{r}_2)
\cdots 
E_{c_m}(\bm{r}_m)
E_{c'_1}^*(\bm{r'}_1) 
E_{c'_2}^*(\bm{r'}_2) 
 \cdots 
E_{c'_{\M}}^*(\bm{r'}_{\M})
\label{eqn:Meas}
\ee
\end{widetext}
where  
$c_i, c'_j \in \{x,y,z\}$ for $i=1,2,\dots,m$,  $j=1,2,\dots,\M$, with $m$ not necessarily equal to $\M$ and 
$\{\bm{r}\} \equiv \{ \bm{r}_1,\cdots \bm{r}_m\},$
$\bm{c} \equiv \{ c_1,\cdots c_m\}.$
The primed variables are denoted similarly. Here $(m,\M)$ denote the order of the electric field correlation tensor.  
We choose to focus on the electric field for the rest of this paper. From Eq. \eref{eqn:SWEpair}, it follows that the magnetic field will carry equivalent signatures.

The correlation tensors are crucial in studying the different orders of optical coherence of the light beam. The theory of partial coherence becomes relevant specifically when investigating the unknown stochastic nature of the intervening medium. In this paper, we study intensity-intensity correlations captured in an interferometer with classical sources of light. If and when captured, the higher-order correlation tensors could potentially reveal more about the nature of the spacetime fluctuations and may lead to stricter bounds.
Another strength of these correlation tensors lies in their natural extendability to quantum sources of light~\cite{ECGSuda,Glaub} such as entangled and squeezed states.
These could also lead to newer and stricter bounds from future experiments using such quantum states~\cite{QLtIRB1,QLtIRB2,prad2020}.

Solving Eq.~\eqref{eqn:SWE} for any $n(\bm{r})$ to obtain the above field correlations in closed form analytically
seems intractable, although numerical solutions may be possible. Seeking a closed form solution of Eq. \eref{eqn:SWE}, we employ the following assumptions motivated by 
an electromagnetic wave propagating along a certain direction, say $z$.

\begin{enumerate}[label=(\roman*)]
\item \label{Ass:VarSep} Separation of variables for the electric field. This relies on Attribute \ref{mod:SmFl} of our model in Sec. \ref{sec:model}. 

\item \label{Ass:Parax} Paraxial and slowly varying envelope approximation (SVEA)~\cite{SVEA93}.  

\end{enumerate}

Using both the ansatz, $E_{y}(x,y,z,t)=\mathcal{S}(x,y,z) \mathcal{T}(t)$ ($i=x,y,z$) and the Attribute \ref{mod:SmFl}, we invoke Assumption \ref{Ass:VarSep}. This separation of variables splits Eq. \eref{eqn:SWE} into the Helmholtz-type equations, 
\begin{eqnarray}
\nabla^{2} \mathcal{S} = - k^{2} \nS^{2} \mathcal{S},
\label{eqn:WEInt1}\\
\frac{\partial^{2} \mathcal{T}}{\partial t^{2}} = - \frac{k^{2} c^{2}}{\nt^{2}} \mathcal{T},
\label{eqn:WEInt2}
\end{eqnarray}
with $k=2 \pi/\lambda=\Omega/c,$. where $\lambda$ is the wavelength of light and $\Omega$  the corresponding frequency. 
The spatial part of the field $\mathcal{S}(x,y,z)$ (respectively, the temporal part, $\mathcal{T}(t)$) is any function that satisfies Eq. \eref{eqn:WEInt1} (respectively, Eq. \eref{eqn:WEInt2}).

Subsequently using $\mathcal{S}(x,y,z)=E_{\textsc{s}}(x,y,z) e^{-i k z}$ and $\mathcal{T}(t)=E_{\textsc{t}}(t) e^{i \Omega t}$, 
and applying \ref{Ass:Parax}, that is, 
\begin{equation}
k \frac{\partial E_{\textsc{s}}}{\partial z} \gg \frac{\partial^{2} E_{\textsc{s}}}{\partial z^{2}}, \quad \mathrm{and} \quad \Omega \frac{\partial E_{\textsc{t}}}{\partial t} \gg \frac{\partial^{2} E_{\textsc{t}}}{\partial t^{2}},
\end{equation}
Eq.~\eqref{eqn:SWE} simplifies to
\begin{subequations}
\be
\frac{\partial E_{\textsc{s}}}{\partial z} = -\frac{i}{2 k} \left( \frac{\partial^{2} E_{\textsc{s}}}{\partial x^{2}} + \frac{\partial^{2} E_{\textsc{s}}}{\partial y^{2}} \right)- \frac{i k}{2} (\nS^{2}-1) E_{\textsc{s}},
\ee
\be
\frac{\partial E_{\textsc{t}}}{\partial t} = - \frac{i \Omega}{2} E_{\textsc{t}} + \frac{i k^{2} c^{2}}{2 \Omega \nt^{2}} E_{\textsc{t}}.
\ee
\label{eqn:SDEs}
\end{subequations}

Now invoking Attribute \ref{mod:small} and neglecting terms of order
$w_{\textsc{x}}^{2}$ ($\textsc{x = s,t}$) and above, the corresponding solutions are
\be
 E_{\textsc{s}}(x,y,z) = \iint_{-\infty}^{\infty} \rmd k_{x} \rmd k_{y} e^{-i (k_{x} x + k_{y} y)}  \widetilde{E}_{\textsc{s}}(k_{x},k_{y},z),
 \label{eqn:SolnS}
\ee
\be
E_{\textsc{t}}(t)=E_{\textsc{t}}(0) \, \exp \left( - i\, \Omega\, \int\limits_{0}^{t} \rmd t' \, \wt(t') \right),
\label{eqn:SolnT}
\ee
where
\begin{eqnarray}
\nonumber \widetilde{E}_{\textsc{s}}(k_{x},k_{y},z)&=&\widetilde{E}_{\textsc{s}}(k_{x},k_{y},0) \, e^{i z \left( k_{x}^{2} + k_{y}^{2} \right)/2 k}\\
&& \hspace*{1.5 em} \exp \left(- i k \int\limits_{0}^{z} \rmd z' \, \ws(x,y,z') \right),
\label{eqn:IFTs}
\end{eqnarray}
and $k_{x}$, $k_{y}$ are the Fourier-conjugate variables corresponding to $x$ and $y$ respectively. 

We note that the spatial component of the propagating field captures the cumulative fluctuations along the direction of traversal 
via $\int\limits_{0}^{z} \rmd z' \, \ws(x,y,z')$ in the phase factor of Eq. \eref{eqn:SolnS}.
This remains unchanged even on implementing the inverse Fourier transform (Eq. \eref{eqn:IFTs}). This effective phase fluctuation with the $x,y$ dependence modifies the shape of the wavefront at each point as the light traverses along the $z$-axis. 
The temporal component also has effective phase fluctuation as in Eq. \eref{eqn:SolnT}. 
As the fluctuations appear in the phase, it follows from Eq. \eref{eqn:SolnT} that some correlation tensors given by Eq.~\eqref{eqn:Meas} do not contain their signatures,
such as $M^{m,m}_{\bm{c}; \bm{c}} \left( \{\bm{r} \}; \{\bm{r} \} \right)$ which is an instantaneous intensity correlation function. 

Others, such as the second-order correlation function 
\be
M^{1,1}_{y;y} \left(\bm{r}_1; \bm{r'}_1 \right)
=
E_y(\bm{r}_1)E^*_y(\bm{r'}_1)
\ee
 do contain signatures of spacetime fluctuations.
Indeed, any measurement that  records some aspect of  the wavefront or reconstructs it entirely should
 reveal observable signatures of the underlying spacetime fluctuations. 

A typical measurement can involve measuring the second-order correlation function $M^{1,1}_{y;y} \left(\bm{r}_1; \bm{r'}_1 \right)$ over a cross-section of a beam at $z=z_{\mathrm{det}}$ and time $t=t_{\mathrm{det}}$, 
such that $\bm{r}_1=(x_{1},y_{1},z_{\mathrm{det}},t_{\mathrm{det}})$ and $\bm{r'}_1=(x_{2},y_{2},z_{\mathrm{det}},t_{\mathrm{det}})$. 
In this case, the temporal part of the correlation function, $E_{\textsc{t}}(t_{\mathrm{det}}) E^{\ast}_{\textsc{t}}(t_{\mathrm{det}})$, does not contain any signatures of spacetime fluctuations. Only the spatial part reveals signatures of interest. 
This is why it was not necessary to specify $\varrho(\delta_t)$ in Attribute~\ref{mod:gauss}. This argument is applicable even when considering the appropriate correlation tensor in the context of interferometric setups in subsequent sections.

There is a great variety of possible configurations of the field correlation tensors in Eq.~\eqref{eqn:Meas} whose measurements 
can reveal signatures of spacetime fluctuations. 
We leave a fuller exploration for the future.
In the rest of this paper, we focus on how interferometric measurements can identify signatures of spacetime fluctuations~\cite{amelino99}.

\begin{figure*}
\includegraphics[width=0.49\textwidth]{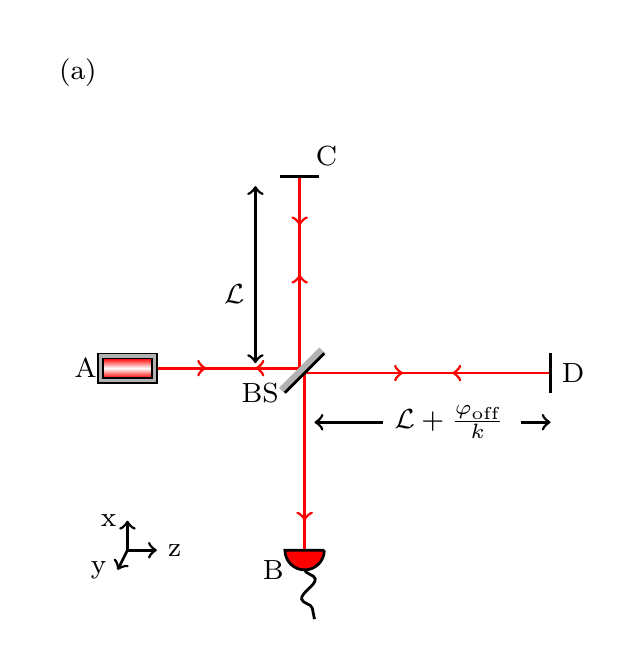}
\includegraphics[width=0.49\textwidth]{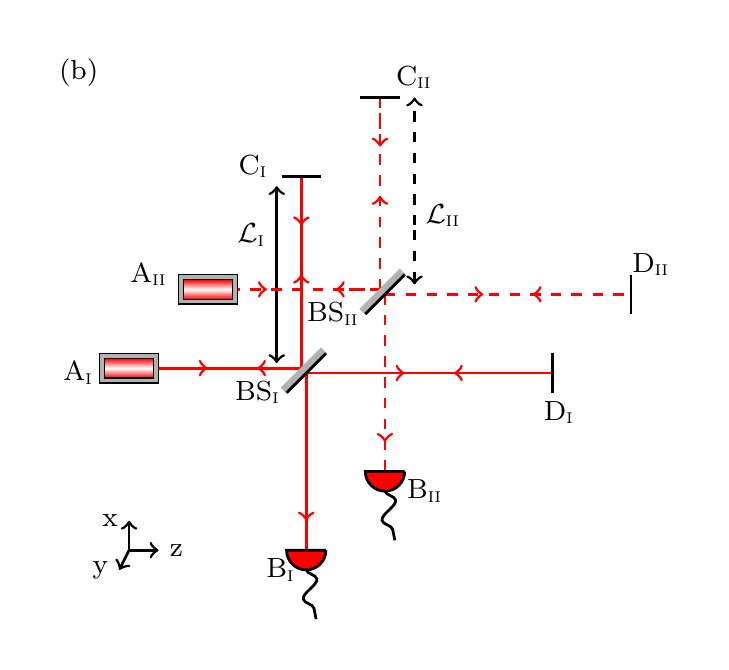}
\caption{A schematic diagram of the interferometric setup. (a) Michelson interferometer with a laser source at the input port A and a detector at the output port B with the two perpendicular arms denoted by C and D. 
The 50/50 lossless beamsplitter is denoted by BS and is taken as the origin of the reference frame in our computation. We can effectively assume the detector to be at the origin as any change suffered by the light after interference at the BS is common to output field contributions from both the arms and therefore cannot be detected. (b) Two co-located Michelson interferometers with input ports A$_{i}$ and output port B$_{i}$ with the two perpendicular arms denoted by C$_{i}$ and D$_{i}$ each with arm length $\Li^{i}$ ($i=$I,II). The origin is at BS$_{\textsc{i}}$.
}
\label{fig:HoloSetup}
\end{figure*}

\vspace*{1 ex}
\section{\label{sec:IntSig} Interferometric Signatures}

Given their extremely high sensitivities, laser interferometers have been proposed and used to probe spacetime fluctuations \cite{amelino99,int1,cardiff21,geontropy2023}. 
As shown in Fig. \ref{fig:HoloSetup} (a), the power of the light at the output port B of such an interferometer, with the beam cross-section in the $y$-$z$ plane, is measured by the photodetector.  
The electric field at the output is given by
\begin{equation}
E_{\mathrm{out}}(\bm{r}_0) = \frac{1}{\sqrt{2}} \bigg[ E^{(\textsc{c})}_{y}(\bm{r}_0) - E^{(\textsc{d})}_{y}(\bm{r}_0) e^{-2 i \varphi_{\text{off}}} \bigg],
\label{eq:Eout}
\end{equation}
where the superscripts $(\textsc{c})$ and $(\textsc{d})$ denote the $y$-component of the electric field along the respective arms of the interferometer,
$\bm{r}_0 \equiv \{0,y,z,\tau_{0} \},$ and $\tau_{0}$ is the time taken to complete a round trip in either of the arms. 
We note that $\varphi_{\mathrm{off}}$ is the phase difference due to the offset in the interferometer arm lengths between the two arms of each interferometer. As the arms are almost of equal length $\Li$ with $\varphi_{\text{off}}/k \ll \Li$, $\tau_{0}=2 \Li / c$. This deliberate small phase offset is present in interferometers that use the so-called DC-readout scheme (including all interferometers considered here), which makes the destructive interference at the output imperfect~\cite{bond2016}.

For brevity, we define the correlation function 
$M_{\mathrm{out}}^{m,\M} \left(\{\bm{r}\}; \{\bm{r'}\} \right)$ for the output field of an interferometer in line with $M^{m,\M}_{\bm{c};\bm{c'}} \left(\{\bm{r}\}; \{\bm{r'}\} \right)$,
except with $E_{\mathrm{out}}$ replacing the field components $E_{c}$ and $E_{c'}$ in Eq. \eref{eqn:Meas}\footnote{Similarly, when different combinations of $ E^{(\textsc{x})}_{y}$ ($\textsc{x=c,d}$) replace the field components $E_{c}$ and $E_{c'}$ in Eq. \eref{eqn:Meas}, a correlation function $M_{\{\textsc{x}\};\{\textsc{x}'\}}^{m,\M} \left(\{\bm{r}\}; \{\bm{r'}\} \right)$ is defined. This is used in Appendix \ref{app:SDcomput}.}.
In particular, the fourth-order correlation function $M_{\mathrm{out}}^{2,2} \left(\bm{R}; \bm{R} \right),$ $\bm{R} \equiv \{\bm{r}_{1},\bm{r}_{2}\}$ with $\bm{r}_{1}=(0,y_{1},z_{1},\tau_{1})$ and $\bm{r}_{2}=(\delta_{s},y_{2},z_{2},\tau_{2})$ will be  of interest both for a single interferometer as in Fig. \ref{fig:HoloSetup} (a) as well as
 two co-located interferometers~\cite{int1,cardiff21} as in Fig. \ref{fig:HoloSetup} (b).

The time-averaged output power of a single interferometer measured at time $\tau$ is then \begin{eqnarray}
P_{\mathrm{out}}(\tau)= \frac{\epsilon_{0} c}{\Delta_{T}}  \int\limits_{\tau}^{\tau+\Delta_T} \hspace*{- 1 em} \mathrm{d}\tau_{1}  \int_{A} 
\mathrm{d}s \, 
M^{1,1}_{\mathrm{out}} \left(\bm{r}_0; \bm{r}_0 \right),
\label{eqn:PfromE}\\
\nonumber 
\end{eqnarray}
where $M^{1,1}_{\mathrm{out}} \left(\bm{r}_0; \bm{r}_0 \right) = |E_{\mathrm{out}}(\bm{r}_0)|^{2},$ $ \mathrm{d}s \equiv \mathrm{d}y \: \mathrm{d}z,$
 $\Delta_{T}$ is integration time of the detector, and $\int_{A}$ denotes a surface integral over the beam cross-section. 
Similarly, the time-averaged power correlation is
\begin{widetext}
\be
P_{\mathrm{out}}^{i}(\tau) P_{\mathrm{out}}^{j}(\tau+\delta_{\tau})
= \left(\frac{\epsilon_{0} c}{\Delta_T} \right)^{2}  \iint\limits_{\tau,\tau+ \delta_{\tau}}^{\tau+\Delta_T,\tau+ \delta_{\tau}+ \Delta_T} \!\!\! \! \mathrm{d}\tau_{1} \: \mathrm{d}\tau_{2} 
	\iint\limits_A  \!\! \mathrm{d}s_{1} \: \mathrm{d}s_{2} \: M^{2,2}_{\mathrm{out}} \left(\bm{R};\bm{R}\right),
\label{eqn:PPfromM22}
\ee
\end{widetext}
where $i,j= \textsc{I,II}$ denote the interferometers which determine the different $E_{\mathrm{out}}$ fields appearing in $M^{2,2}_{\mathrm{out}} \left(\bm{R};\bm{R}\right).$ 
An explicit expression is given in Appendix~\ref{app:SDcomput}, Eq.~\eqref{eqn:m22DC}.

In interferometric setups such as the Holometer, the control systems carrying out measurements monitor fluctuations in the output light 
and feed back the differential signals to actuated end mirror mounts aimed to hold the average output power constant~\cite{int1}. 
The measurement of relevance is thus a time-series of the deviation from the average output power
accumulated over periods of $T_{\text{acc}} = 1.8/700~\mathrm{s} = 2.6$ ms~\cite{int1}.
Within each $T_{\text{acc}},$ the average output power itself is obtained by averaging over $N_{\text{sp}}=T_{\text{acc}}/\Delta_{T}$ time measurements.

Equations \eqref{eqn:PfromE} and \eqref{eqn:PPfromM22} 
capture the experimental scenario where $n(x,y,z,t)$ (equivalently, $\ws$ and $\wt$) is sampled as the light propagates in space.
However, our theoretical formalism requires assuming specific characteristics of the distribution of $n(x,y,z,t)$ to obtain analytical expressions.
 While the former can be used to obtain the necessary temporal averages, the latter can only be used to obtain ensemble averages.
We thus use the stationarity of the time-series of the deviation from the average output power, to replace time averages with ensemble averages.
This can be justified by the tests establishing the stationarity of the Holometer/aLIGO data which is necessary for the ergodicity of a process.

As the recorded data from experiments such as the Holometer are the power spectral density (PSD) from a single interferometer's output 
or the cross-spectral density (CSD) from the outputs of two co-located interferometers~\cite{LIGOdata,int1,cardiff21},
we invoke ergodicity for the correlation functions of the fourth-order in the electric fields, leading to
\be
\overline{P_{\mathrm{out}}^{i}(\tau) P_{\mathrm{out}}^{j}(\tau+\delta_{\tau})}
= \left(\epsilon_{0} c \right)^{2} \iint\limits_A  \!\! \mathrm{d}s_{1} \: \mathrm{d}s_{2} \: \overline{M^{2,2}_{\mathrm{out}} \left(\bm{R};\bm{R}\right)},
\label{eqn:P2ptCorr}
\ee
and the covariance
\begin{eqnarray}
\nonumber &&\mathrm{Cov}_{i,j}(P_{\mathrm{out}})=\overline{P^{i}_{\mathrm{out}}(\tau) P^{j}_{\mathrm{out}}(\tau+\delta_{\tau})}\\
&& \hspace*{8 em}  \, -\overline{P^{i}_{\mathrm{out}}(\tau)} \:\:\: \overline{P^{j}_{\mathrm{out}}(\tau+\delta_{\tau})}.
\label{eqn:Covij}
\end{eqnarray}
Note that $\mathrm{Cov}_{i,j}(P_{\mathrm{out}}) \neq \mathrm{Cov}_{j,i}(P_{\mathrm{out}})$, 
in general\footnote{In the case $\Li_{\textsc{i}}=\Li_{\textsc{ii}}$, it is seen that $\mathrm{Cov}_{i,j}(P_{\mathrm{out}}) = \mathrm{Cov}_{j,i}(P_{\mathrm{out}})$ in Appendix \ref{app:SDcomput}.}, for $i \neq j$.
Extending ergodicity  to higher-order moments will require  commensurate tests of stationarity on corresponding data.

The PSD is thus
\begin{equation}
S(f)=\frac{1}{\pi} \int_{0}^{\infty} \rmd \delta_{\tau} \, \mathrm{Cov}(\Delta x) \cos 2 \pi f \delta_{\tau}.
\label{eqn:WienKhin}
\end{equation}
and the CSD is
\begin{eqnarray}
\nonumber CS(f)= &\dfrac{1}{2 \pi} \mathop{\mathlarger{\int}}_{\hspace*{-0.5 em}0}^{\infty} \rmd \delta_{\tau} \, \left\{ \rule{0cm}{0.55cm}\right. e^{-2 \pi i f \delta_{\tau}}  \, \mathrm{Cov}_{\textsc{i,ii}}(\Delta x) \\
&+  e^{2 \pi i f \delta_{\tau}}  \, \mathrm{Cov}_{\textsc{ii,i}}(\Delta x) \left. \rule{0cm}{0.55cm} \right\},
\label{eqn:CSdefn}
\end{eqnarray}
where\footnote{The average output power $P_{\mathrm{out}}(\tau)$ is scaled by 
$(2 k \varphi_{\mathrm{off}} P_{0})^{-1}$
to get the corresponding path difference, where $P_{0}$ is the power of the input laser beam. This scaling factor is obtained using the following argument.
Given that there is an effective phase difference $\varphi_{s}$ between the two arms of an interferometer, the average output power can also be written as $P_{\mathrm{out}}(\tau)=P_{0} \sin^{2}(\varphi_{s}+ \varphi_{\mathrm{off}})$. Expanding this to first order in $\varphi_{s}$ and $\varphi_{\mathrm{off}}$, and denoting the corresponding path difference by $\delta \Li$, the effective phase difference is
\begin{equation}
\nonumber
\varphi_{s} = k \, \delta \Li \approx \frac{P_{\mathrm{out}}(\tau)}{2 \varphi_{\mathrm{off}} P_{0} }.
\end{equation}}
\be
\mathrm{Cov}_{i,j}(\Delta x)= \left(\frac{\lambda}{4 \pi \varphi_{\mathrm{off}} P_{0}}\right)^{2} \mathrm{Cov}_{i,j}(P_{\mathrm{out}}).
\label{eqn:CovXDefn}
\ee
For brevity, $\mathrm{Cov}(\Delta x) \equiv \mathrm{Cov}_{i,i}(\Delta x)$.

In practice, the PSD and the CSD are obtained using discrete Fourier transforms of time series of the measured path difference.
This corresponds to using discretized transforms in place of Eqs.~\eref{eqn:WienKhin} and \eref{eqn:CSdefn}. 
In the next section, we compare the PSDs and CSDs obtained analytically by evaluating these discretized versions of Eqs. \eref{eqn:WienKhin} and \eref{eqn:CSdefn} with the corresponding measured spectral densities. 
These analytical evaluations use Attributes~\ref{mod:stat}, \ref{mod:indep}, and \ref{mod:gauss}
to provide constraints on the strength ($\Gs$) and scale ($N_{\parallel}$) of the spacetime fluctuations.

\section{\label{sec:SDs}Constraint from Spectral Densities}

In the Holometer and QUEST experiments, the setup is a simple Michelson interferometer as represented in Fig. \ref{fig:HoloSetup} (b). 

We show how constraints on the strength and scale of the spacetime fluctuations could be established from the PSD
and then the CSD, both from completed and proposed experiments such as the Holometer~\cite{HoloData}, aLIGO~\cite{LIGOdata}, and QUEST~\cite{cardiff21}. These results rely on Attributes~\ref{mod:stat}, \ref{mod:indep}, and \ref{mod:gauss},
and assume the spacetime fluctuations to be isotropic in space with the scale $N_{\parallel} \equiv N_{x} = N_{y}= N_{z}$. We present the constraints as two-dimensional $\Gs$-$N_{\parallel}$ exclusion plots. Importantly, these constraints only apply when
spectral density produced by the spacetime fluctuations is present at a frequency at which the interferometer is sensitive. As explained below, this is not the case for our model from Sec.~\ref{sec:model} and presently operational interferometers.

\subsection{\label{sec:PSD}Constraint from PSD}

Constraints based on the Holometer and the QUEST experiments are presented in the first two subsections. 
The former is based on experimental data~\cite[Fig. 2]{HoloData}, 
and the latter on proposed experimental parameters~\cite{cardiff21}.
The third subsection reports similar constraints from aLIGO data, which is a Michelson interferometer with cavities in each arm which requires some additional analysis.

\subsubsection*{Holometer}

Using Eq. \eref{eqn:Cov} from Appendix~\ref{app:SDcomput} in the expression for $\mathrm{Cov}_{i,j}(\Delta x)$ in Eq.~\eref{eqn:CovXDefn}, and setting $\delta_{s}=0$ in the discretized version of Eq.~\eref{eqn:WienKhin}, the PSD is
\be
S(f)=\mathcal{A} \left[ 1 + \, \mathcal{B}^2  \cos 4 \varphi_{\mathrm{off}} - 2 \mathcal{B} \cos^{2} 2 \varphi_{\mathrm{off}} \right],
\label{eqn:PSDholocomp}
\ee
where
\begin{eqnarray}
&&\mathcal{A}= \frac{1}{4 \pi}\left(\frac{\lambda}{8 \pi \varphi_{\mathrm{off}}}\right)^{2} \, \Theta(f),\\
&& \mathcal{B} = \exp{\left(- 2 \sqrt{2 \pi} \mathcal{K} \Li \right)},\\
&&\mathcal{K}=k^{2} \ell_{\textsc{p}} \Gs  N_{\parallel},
\end{eqnarray}
and
the step function $\Theta(f)$ (in units of time) is
\begin{equation}
\Theta(f)=\begin{cases}
1 \,\text{s}, & \mathrm{if }\:\: 0 \leqslant f < f_{\mathrm{step}},\\
0 \,\text{s}, & \mathrm{otherwise,}
\end{cases}
\label{eqn:step}
\end{equation}
where $f_{\mathrm{step}}$ denotes the frequency resolution of the discrete Fourier transform, i.e., the width of the frequency bins. 
Thus the signal has all its power in the zero frequency (DC) bin.

To first-order in Planck length, i.e., $\mathcal{B}=1- 2 \sqrt{2 \pi} \mathcal{K} \Li$, the PSD becomes
\begin{equation}
S(f) \approx \left(\frac{\lambda}{2 \pi} \right)^{2} \frac{\mathcal{K} \Li}{2 \sqrt{2 \pi}} \Theta(f).
\label{eqn:PSDholo}
\end{equation}
Here we have used the fact that $\varphi_{\mathrm{off}}\ll 1$, as in the setups considered $\varphi_{\mathrm{off}}\approx 10^{-5}$ rad (see Table \ref{tab:param}).

In our drastically simplified spacetime fluctuations model, the use of Attribute \ref{mod:SmFl} and Assumption \ref{Ass:VarSep} results in an
$M^{2,2}_{\mathrm{out}} \left(\bm{R};\bm{R}\right)$ independent of time due to Eq.~\eref{eqn:SolnT}.
This in turn leads to a covariance that is independent of $\delta_{\tau}$ and thereby, a PSD with a Dirac delta function of the frequency centred on $f=0$. For a discrete cosine transform, $\delta(f)$ is replaced by $\Theta(f).$ This predicted frequency dependence is very different from spectra theorised in other models for quantum spacetime fluctuations \cite{LiZurek2023,kwon2022}. 

Interferometers have limited or no sensitivity to signals at zero frequency. Therefore, a direct comparison between our predicted signal and interferometric observations is not possible, and our model in Sec.~\ref{sec:model} cannot be constrained by existing observations. Nevertheless, to illustrate the principle, we proceed to constrain our model by comparing the order of magnitude of the predicted signal spectral density and the noise of the interferometers, anticipating the possibility of the signal having a different frequency dependence but a similar magnitude.

The sensitive band of the Holometer data (i.e., at frequencies above $f_\mathrm{sens}\approx1$~MHz is limited by photon shot noise~\cite{HoloData}
-- a white noise spectrum in principle~\cite{NoneqStat}, which is given by~\cite{cardiff21}
\begin{equation}
S_{\mathrm{meas}}(f>1\;\mathrm{MHz}) = \frac{\hbar c \lambda}{8 \pi P_{0}}.
\end{equation}

We now consider spacetime fluctuations described by our model to produce a displacement spectral density at frequencies within this band but with a magnitude smaller than that of the shot noise. The model can thus be constrained at a signal-to-noise ratio of one\footnote{We choose to set constraints at SNR=1 because it is conventional in the field of laser interferometry to define the sensitivity of an experiment as the magnitude of a potential signal equal to the instrument's noise. Though the choice is arbitrary, it could straightforwardly be converted to any desired confidence level.} by setting $S_{\mathrm{meas}}(f_\mathrm{sens}>0) \geqslant S(f=0)$, which gives
\begin{equation}
\Gs  N_{\parallel} \leqslant  \frac{\hbar c }{2\sqrt{2 \pi} \, \ell_{\textsc{p}}} \: \frac{\lambda}{P_{0} \Li},
\label{eqn:GNbound}
\end{equation}
We use Eq. \eref{eqn:GNbound} and the parameter values listed in Table \ref{tab:param} in the case of the Holometer, to obtain the allowed parameter space 
as shown in Fig. \ref{fig:ParamSpace} (a).

\begin{table}
\begin{tabular}{lccc}
\hline
\hline
Parameter & Holometer~\cite{HoloData} & QUEST~\cite{cardiff21} & aLIGO~\cite{LIGOdata} \\
\hline
\hline
$P_{0}$ (kW) & 2 & 10 & 750 \\
$\Li$ (m) & 40 & 3 & $4 \times 10^{3}$\\
$\Delta_{T}$ (s) & $10^{-8}$ & $10^{-9}$ & $10^{-5}$\\
$\varphi_{\mathrm{off}}$ (rad) & $5\times 10^{-5}$ & $5\times 10^{-5}$ & $2.5 \times 10^{-5}$\\
$\lambda$ ($\times 10^{-6}$ m) & 1.064 & 1.064 & 1.064 \\
PSD (m$^{2}/$Hz) & $6.3\times 10^{-37}$ & $3.2\times 10^{-38}$ & $2.3\times 10^{-40}$ \\
\hline
\end{tabular}
\caption{\label{tab:param} 
Values of the system parameters as measured in the Holometer~\cite{HoloData} and aLIGO~\cite{LIGOdata}. Values of $\Delta_{T}$ and $\varphi_{\mathrm{off}}$ are up to an order-of-magnitude. System parameter values used in the simulation of the QUEST experiment~\cite{cardiff21} are also quoted.}
\end{table}

Furthermore, \eref{eqn:GNbound} and the subsequent inequalities are valid only between the limits $N_{\parallel} \ll 2 \Li/\ell_{\textsc{p}}$ and  $N_{\parallel} \gg \lambda/\ell_{\textsc{p}}$. We note that the latter limit arises from Eq. \eref{eqn:fluctIneq} as mentioned in Sec. \ref{sec:EMsig}. 
We consider the case $N_{\parallel}=2 \Li/\ell_{\textsc{p}}$ in Appendix \ref{app:LiCorr} for the Holometer. While a constraint on $\Gs$ is obtained in this case too, the product $\Gs  N_{\parallel} \leqslant 10^{-1}$. This is a weaker constraint than $\Gs  N_{\parallel} \leqslant 5 \times 10^{-3}$ obtained using Eq. \eref{eqn:GNbound} at $N_{\parallel}\ll 2 \Li/\ell_{\textsc{p}}$. On the other hand, the constraint from CSD becomes stronger than the corresponding constraint for $N_{\parallel}\ll 2 \Li/\ell_{\textsc{p}}$.

\subsubsection*{QUEST}

The simulation for QUEST considered a shot-noise limited sensitivity above $f_\mathrm{sens}\approx1$~MHz, and the use of squeezed vacuum states of light, aiming to attain 6 dB of squeezing in the output~\cite{cardiff21}. This leads to a measured PSD $S_{\textsc{quest}}(f)$, which is written in terms of $S_{\mathrm{meas}}(f)$ as,
\begin{equation}
S_{\textsc{quest}}(f>1\;\mathrm{MHz}) = 10^{-0.6} S_{\mathrm{meas}}(f>1\;\mathrm{MHz}).
\label{eqn:PSDqst}
\end{equation}
As before, the model can thus be constrained at a signal-to-noise ratio of one by setting $S_{\mathrm{meas}}(f_\mathrm{sens}>0) \geqslant S(f=0)$
\begin{equation}
\Gs  N_{\parallel} \leqslant  10^{-0.6} \, \frac{\hbar c }{2\sqrt{2 \pi} \, \ell_{\textsc{p}}} \: \frac{\lambda}{P_{0} \Li}.
\label{eq:GNqst}
\end{equation}
Using parameters listed in Table \ref{tab:param}, Eq. \eref{eq:GNqst} yields the maximum limit on the product $\Gs  N_{\parallel}$ 
which is plotted in Fig. \ref{fig:ParamSpace} (a) as a line, indicating that an experimental run of the QUEST experiment could rule out the parameter space above this line.
This constraint from the simulation is stricter than that from the Holometer when 6 dB of squeezing is used to achieve sub-shot-noise performance.

\begin{figure}[t!]
\includegraphics[width=0.4\textwidth]{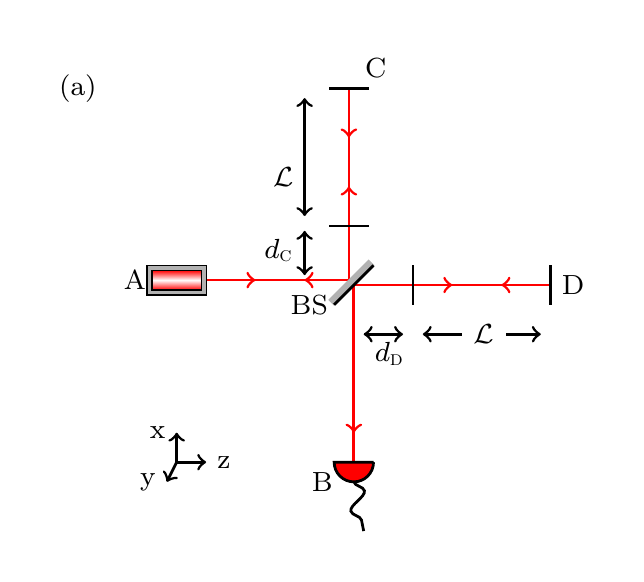}
\includegraphics[width=0.4\textwidth]{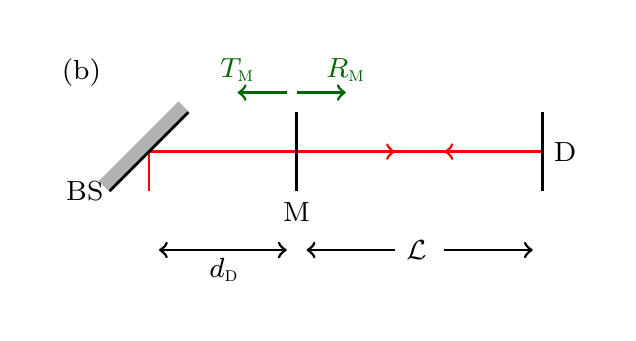}
\caption{(a) Michelson interferometer with arm cavities and $d_{\textsc{d}}-d_{\textsc{c}}=\varphi_{\text{off}}/k$, and (b) arm D of the interferometer.}
\label{fig:LIGO}
\end{figure}

\subsubsection*{aLIGO}

aLIGO is a Michelson interferometer with cavities in each arm~\cite{LIGOdata} as shown in Fig. \ref{fig:LIGO}. These arm cavities are formed by introducing a mirror in each arm.
The electric field at the detector B is 
\begin{eqnarray}
\nonumber &E_{\mathrm{out}}(0,y,z,\tau) = \frac{1}{\sqrt{2}} \sum_{q=1}^{N_{\mathrm{rt}}(\tau)} \sqrt{T_{\textsc{m}} R_{\textsc{m}}^{q-1}}\\ 
&\bigg[ E^{(\textsc{c})}_{y}(d_{\textsc{c}},y,z,\tau) - E^{(\textsc{d})}_{y}(d_{\textsc{d}},y,z,\tau) e^{- 2 i \varphi_{\text{off}}} \bigg].
\label{eqn:EoutLIGO}
\end{eqnarray} 
Here $N_{\mathrm{rt}}(\tau)=\lfloor\frac{c \tau}{2 \Li}\rfloor$ and $T_{\textsc{m}}=1-R_{\textsc{m}}$ 
is the transmission coefficient of the mirrors introduced to render arm cavities.
The analogous expression without arm cavities corresponds to Eq.~\eqref{eq:Eout}.

The PSD to first-order in Planck length is
\begin{equation}
S(f) \approx \frac{\mathcal{K} \Li}{2 \sqrt{2 \pi}} \left(\frac{\lambda T_{\textsc{m}}}{2 \pi}\right)^{2}  \,  \left(\frac{1- (\sqrt{R_{\textsc{m}}})^{280}}{1- \sqrt{R_{\textsc{m}}}}\right)^{4} \Theta(f).
\label{eqn:PSDligo}
\end{equation}
The derivation is in Appendix \ref{app:aLIGOcov}.  
We use $N_{\mathrm{rt}}(\tau)=280$ based on $T_{\textsc{m}}=0.014,$ whereby $R_{\textsc{m}}^{280} \approx 0.019 < 0.02$, 
i.e., less than 2$\%$ of the input light remains after $280$ round-trips of the light beam in each arm.

\begin{figure*}
\includegraphics[width=0.45\textwidth]{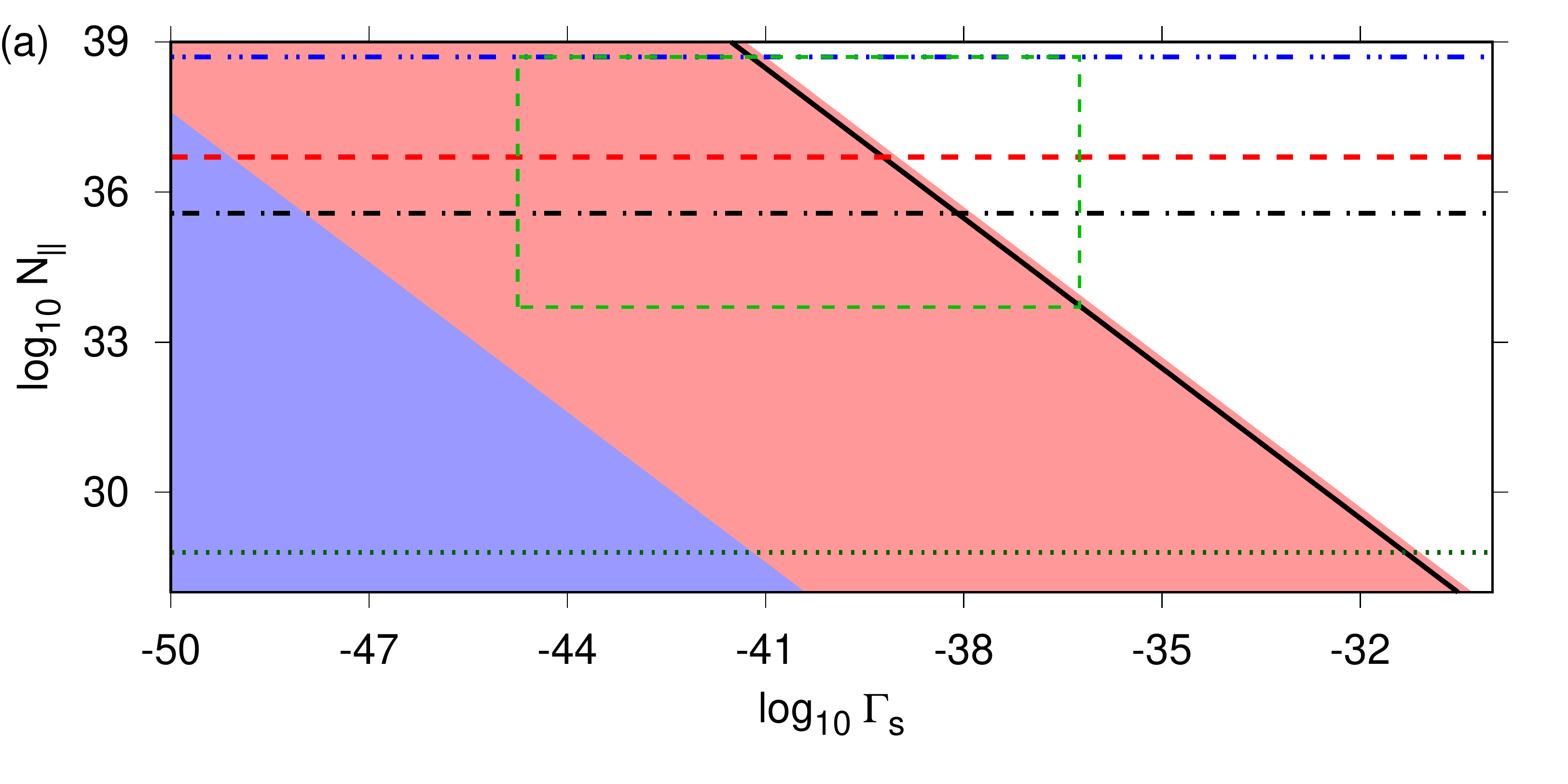}
\includegraphics[width=0.45\textwidth]{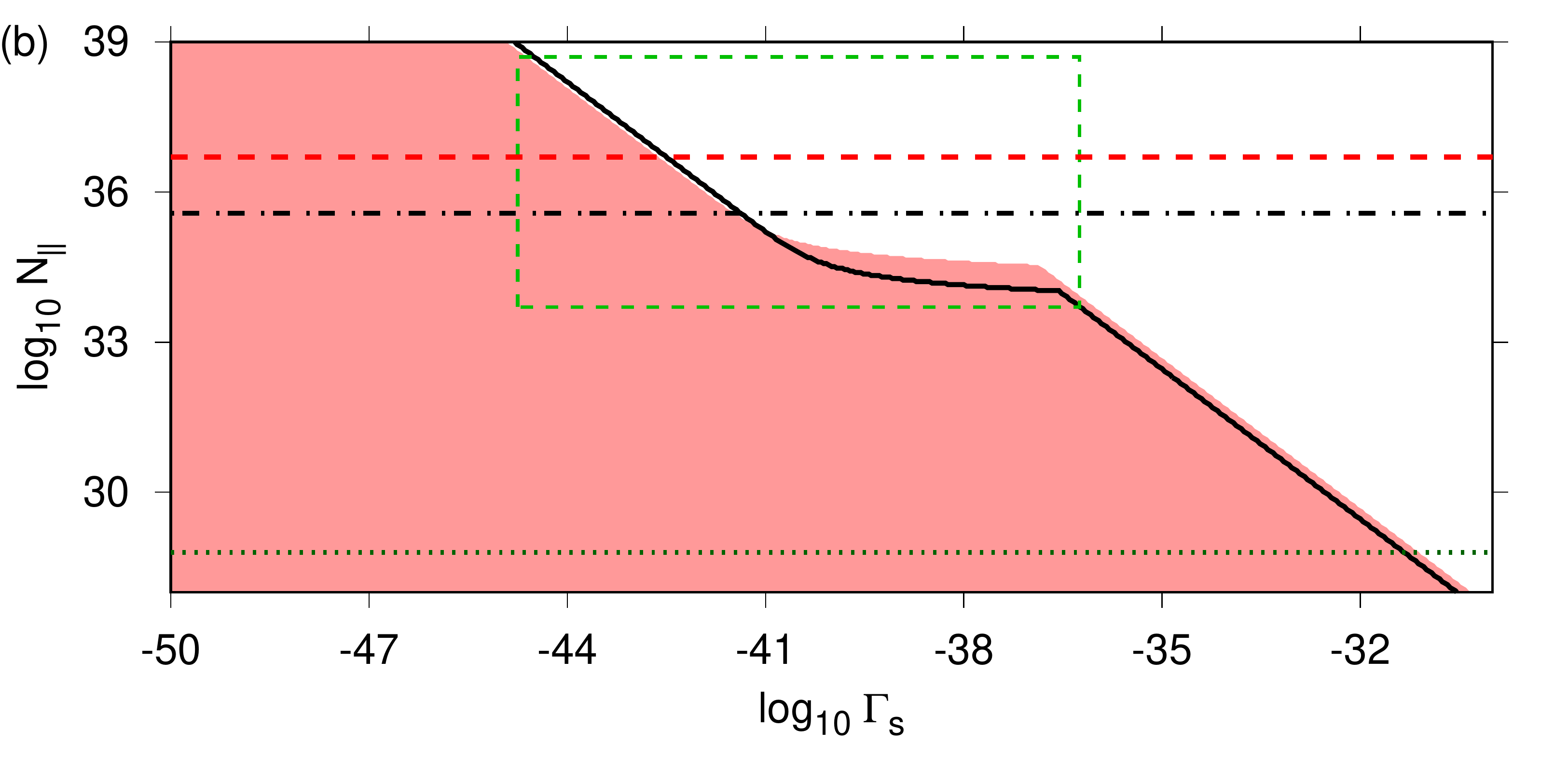}
\caption{(a) The shaded area demarcates the allowed parameter space for $\Gs$ and $N_{\parallel}$ as predicted from the measured PSD in the Holometer (red+blue) and in the aLIGO setup (blue). Comparing $S(f)$ to the simulation of the QUEST experiment gives an upper bound on $\Gs N_{\parallel}$ in the case of 6 dB of squeezing (black solid line). 
(b) The shaded area demarcates the allowed parameter space for $\Gs$ and $N_{\parallel}$ as predicted from the measured CSD in the Holometer (red). The solid black line gives an upper bound on the allowed parameter values of $\Gs$ and $N_{\parallel}$ using the condition $|CS(f=0)|\leqslant |CS_\mathrm{meas}(f>f_\mathrm{sens})|= 10^{-41} $ m$^{2}$ Hz$^{-1}$ with $f_\mathrm{sens}=1$~MHz in the case of the QUEST experiment with 6 dB of squeezing.
In both (a) and (b), the horizontal lines correspond to $N_{\parallel} = 2 \Li/\ell_{\textsc{p}}$ in the Holometer (dashed, red), aLIGO (dot-dot-dashed, blue), and the QUEST (dot-dashed, black) experiments respectively. 
The dark green horizontal dotted line corresponds to $N_{\parallel} = \lambda/\ell_{\textsc{p}}$. The light green dashed rectangle marks the parameter space in which we see a significant increase in information from CSD measurement.}
\label{fig:ParamSpace}
\end{figure*}

Again, we compare $S(f)$ in Eq.~\eref{eqn:PSDligo} with the experimentally measured PSD at the frequency of maximum sensitivity ($f_\mathrm{sens}=235$~Hz, attained at the LIGO Hanford site)~\cite{aLIGOdata}), and we constrain the product $\Gs N_{\parallel}$ as in Fig. \ref{fig:ParamSpace} (a).

\subsection{\label{sec:CS}Constraint from CSD}

The cross-correlation between the output power from the two co-located interferometers is captured using the cross-spectral density $CS(f)$ in Eq. \eref{eqn:CSdefn}. 

We first compute the $CS(f)$ in the case of the Holometer, where the beamsplitters of the two co-located interferometers are separated by 0.91 m diagonally in 3 dimensions with $\delta_{s}=0.91/\sqrt{3}$ m being the separation along each axis. We consider the two interferometers to have equal arm lengths $\Li_{\textsc{i}}=\Li_{\textsc{ii}}=\Li$. 
Defining 
\begin{eqnarray}
&&\mathcal{D}=\exp \left(-\frac{4 \delta_{s}^{2}}{ \ell_{\textsc{p}}^{2} N_{\parallel}^{2}}\right),
\label{eqn:Ddefn}
\end{eqnarray}
and using Eqs. \eref{eqn:CSdefn}, \eref{eqn:CovXDefn}, and \eref{eqn:Cov}, we obtain the cross-spectral density, as
\be
CS(f)= \mathcal{A}  \left[\mathcal{B}^{1-\mathcal{D}} + \mathcal{B}^{1+\mathcal{D}}  \, \cos 4 \varphi_{\mathrm{off}}  - 2  \mathcal{B} \cos^{2} 2 \varphi_{\mathrm{off}} \right].
\label{eqn:CSeval}
\ee
As $\mathcal{D} \rightarrow 0, CS(f) \rightarrow 0$, and as $\mathcal{D} \rightarrow 1, CS(f) \rightarrow S(f)$
in Eq.~\eref{eqn:PSDholocomp}.
For intermediate values of $\mathcal{D}$, the CSD thus provides additional information beyond the PSD. 
It adds to the constraint on $\Gs  N_{\parallel}$ such as in Eqs.~\eref{eqn:GNbound} or \eref{eq:GNqst} 
which appears as a straight line of the log-log plot in Fig. \ref{fig:ParamSpace} (a).
This shows the additional strength of our two-parameter model as compared to a single-parameter one capturing only $\Gs  N_{\parallel}$.

As with the PSD, we use the measured cross-spectral density 
in the frequency range $f > f_{\mathrm{sens}}$, denoted by $|CS_\mathrm{meas}(f>f_\mathrm{sens})|$, to set the upper bound on $|CS(f)|$ at $f=0$.
For the Holometer with $|CS_\mathrm{meas}(f>f_\mathrm{sens})| =10^{-40} $ m$^{2}$ Hz$^{-1}$~\cite{HoloData} , the constraint is shown in Fig. \ref{fig:ParamSpace} (b). 
Comparing Figs. \ref{fig:ParamSpace} (a) and (b), it is evident that considering the cross-correlations has improved the bound on $\Gs$ and $N_{\parallel}$. We see that the constraint is improved only where $\ell_{\textsc{p}} N_{\parallel}$ is comparable to $\delta_{s}.$
It is thus evident that stricter constraints can be obtained by reducing the separation between the co-located interferometers.

In the proposed QUEST experiment, the beamsplitters are to be separated by 0.3 m along the diagonal \textit{in the plane} of the interferometer~\cite{cardiff21}. 
Using Eq. \eref{eqn:CSeval} with $\delta_{s}=0.3/2$ m  and an estimated $|CS_\mathrm{meas}(f>f_\mathrm{sens})| = 10^{-41} $ m$^{2}$ Hz$^{-1}$ 
for 6 dB of squeezing we obtain the improved bound on $\Gs$ and $N_{\parallel}$ displayed in Fig. \ref{fig:ParamSpace}(b). 
The choice of $\delta_{s}=0.3/2$ m in place of $\delta'_{s}=0.3/\sqrt{2}$ m is explained in Appendix \ref{app:SDcomput} using Eq. \eref{eqn:genDdefn}, 
while $|CS_\mathrm{meas}(f>f_\mathrm{sens})| = 10^{-41} $ m$^{2}$ Hz$^{-1}$ is obtained by combining
(A) shot-noise-limited PSD $S_{\textsc{quest}}(f)\approx 10^{-38}$  m$^{2}$ Hz$^{-1}$ for 6 dB of squeezing (evaluating Eq. \eref{eqn:PSDqst}),
(B) an assumed total measurement time of $T_{\mathrm{tot}}=10^{6}$ s, and
(C) $CS_{\mathrm{meas}}(f)$ is of the order-of-magnitude of $S_{\mathrm{meas}}(f) / \sqrt{T_{\mathrm{tot}}}$.

In the case of aLIGO, the two interferometers are located $\delta_{L} \approx 3030.13$ km apart. The corresponding decay factor $\mathcal{D}$ decreases exponentially and vanishes as $\frac{\delta_{L}}{\ell_{\textsc{p}} N_{\parallel}}\gg 1$ given that $N_{\parallel}\ll \Li/\ell_{\textsc{p}}$. Therefore the limits are set only by the PSD measurements.

\section{\label{sec:alphaComp}Comparison with alternative models}

Our model of spacetime fluctuations and the constraints placed upon it by experimental data can 
be translated to some prior models of spacetime fluctuations. 
One such class is captured by~\cite{AlphMod}
\begin{equation}
\frac{\Delta x}{\Li} \geqslant \left(\frac{\ell_{\textsc{p}}}{\Li}\right)^{\alpha},
\label{eq:alpha}
\end{equation}
where $\Delta x$ is the standard deviation of the path difference in an interferometric setup due to spacetime fluctuations. From Eqs.~\eref{eqn:PSDholo} and \eref{eqn:PSDligo}, it is evident that the PSD is directly proportional to $\mathcal{K}$ and hence, to the product $\Gs N_{\parallel}$ in all three interferometric setups considered. Therefore, we introduce the combined parameter $\Pi=\Gs N_{\parallel}$ to relate our two-parameter approach to the single-parameter $\alpha$ model. Comparing $\Delta x$ with the corresponding  measured $(\Delta x)_{\mathrm{meas}}^{2}$ due to noise, given by
 \begin{align}
  (\Delta x)_{\mathrm{meas}}^{2} &= 4 \pi \int\limits_{0}^{\infty} \rmd f \left[\left.S(f)\rule{0cm}{0.4 cm} \right\vert_{\Pi_{\mathrm{max}} \equiv \text{bound}(\Pi)}  \right],
 \label{eqn:revWKT}
 \end{align}
provides a lower bound on $\alpha.$ 
Here, $\text{bound}(\Pi)$ denotes the value of $\Pi$ such that it gives the strictest bound $\Gs N_{\parallel}\leqslant \Pi_{\text{max}}$, from $\Pi$ computed at the boundary of any allowed parameter space. For instance, in the context of PSD measurements, $\Pi_{\mathrm{max}}$ is the value of $\Pi$ corresponding to the straight-line boundary of the allowed parameter space in Fig.~\ref{fig:ParamSpace}(a). As these allowed parameter spaces are themselves bounded using the measured data, Eq.~\ref{eqn:revWKT} is valid and along with Eq.~\eref{eq:alpha}, it provides the link between $\Pi_{\text{max}}$ and $\alpha$.

For the Holometer, Eq.~\eref{eqn:PSDholo} gives
\begin{equation}
(\Delta x)_{\mathrm{meas}}^{2} = \sqrt{\frac{\pi}{2}} \, \Pi_{\mathrm{max}} \ell_{\textsc{p}} \Li.
\label{eqn:GsLBound}
\end{equation}
Combining this with Eq.~\eref{eq:alpha} gives
\begin{equation}
\alpha \geqslant \frac{1}{2} - \frac{\log \left( \sqrt{\pi/2} \, \Pi_{\mathrm{max}} \right)}{2 \log \left( \Li/\ell_{\textsc{p}} \right)}.
\label{eqn:compBoundHolo}
\end{equation}
Similarly, using Eq.~\eqref{eqn:PSDligo} for aLIGO,
\begin{equation}
\alpha \geqslant \frac{1}{2} - \frac{\log \left( \sqrt{\pi/2} \, \Pi_{\mathrm{max}} \, T_{\textsc{m}}^{2} \: \left(\frac{1- (\sqrt{R_{\textsc{m}}})^{280}}{1- \sqrt{R_{\textsc{m}}}}\right)^{4}\right)}{2 \log \left( \Li/\ell_{\textsc{p}} \right)}.
\label{eqn:compBoundLigo}
\end{equation}

Lower bounds on the value of $\alpha$ can thus be obtained from our constraints on $\Gs N_{\parallel} \leqslant \Pi_{\mathrm{max}}$ obtained for different interferometric setups, such as in Eqs.~\eref{eqn:GNbound} or \eref{eq:GNqst}. 
We present the numerical estimates in Table \ref{tab:comp}.

\begin{table}
\begin{tabular}{llccc}
\hline
\hline
& Setup & Constraint & $\Pi_{\mathrm{max}}$ & $\alpha_{\mathrm{min}}$ \\
& &  from & & \\
\hline
\hline
\multirow{5}{*}{\rotatebox[origin=c]{90}{Interferometric}}&\multirow{2}{*}{Holometer} & PSD & $5 \times 10^{-3}$ & 0.53\\
\vspace*{1 ex}
&& CSD & $1 \times 10^{-6}$ & 0.58\\
\vspace*{1 ex}
&\multirow{2}{*}{QUEST} & PSD & $3 \times 10^{-3}$ & 0.53\\
\vspace*{1 ex}
&& CSD & $1 \times 10^{-6}$ & 0.58\\
\vspace*{1 ex}
&aLIGO & PSD & $4 \times 10^{-13}$ & 0.60\\
\hline
\vspace*{1 ex}
\multirow{4}{*}{\rotatebox[origin=c]{90}{Astronomical}}& Optical &\cite{NgPerlman2022} & -- & 0.53 \\
\vspace*{1 ex}
& X-ray &\cite{NgPerlman2022} & -- & 0.58 \\
\vspace*{1 ex}
& GeV $\gamma$-ray &\cite{NgPerlman2022} & -- & 0.67 \\
\vspace*{1 ex}
& TeV $\gamma$-ray  &\cite{NgPerlman2022} & --  & 0.72 \\
\hline
\end{tabular}
\caption{\label{tab:comp}Comparison of $\Pi_{\mathrm{max}}$ and $\alpha_{\mathrm{min}}$. 
For completeness, we also quote the $\alpha_{\mathrm{min}}$ values obtained from the astronomical observations~\cite[Table 1]{NgPerlman2022}.}
\end{table}

Stronger lower bounds can be placed on $\alpha$ from CSD measurements. 
This is done by taking the maximum value of $N_{\parallel}$ allowed for different allowed values of $\Gs$ to compute
 a set of values $\Gs N_{\parallel}$; the constraint is then set at $\Pi_{\mathrm{max}}$, which is the minimum value in that set. The corresponding $\alpha_{\mathrm{min}}$ is computed using Eq. \eref{eqn:compBoundHolo} and reported in Table \ref{tab:comp}. We have also presented the $\alpha_{\mathrm{min}}$ values obtained from the astronomical observations~\cite{Perlman2015,NgPerlman2022}. For instance, in \cite{Perlman2015}, spacetime fluctuations are shown to inhibit astronomical observations at sufficiently short wavelengths. This allows using X-ray and $\gamma$-ray observations to place stricter constraints as evident from the larger $\alpha_{\mathrm{min}}$. It is interesting to note that the $\alpha_{\mathrm{min}}$ obtained using the CSD measurements in our approach place stricter bounds than that obtained from optical astronomical observations.

Our model is not directly comparable to that by Zurek and coworkers~\cite{VerZurek21,Zurek2022,geontropy2023}.
This is because of the different assumptions in the two models of fluctuations. 
For instance, in Ref.~\cite{geontropy2023}, the metric itself is identified with a scalar field satisfying the massless free scalar wave equation. 
This leads to correlations, particularly angular ones, in the spectral densities that are absent in our model.
Our Attribute \ref{mod:indep} removes possible correlations in the spatial and temporal fluctuations,
 enables the separation of variables in Assumption \ref{Ass:VarSep}, and thereby leads to spectral densities that are delta functions.

\section{Conclusion}
\label{sec:disc}

We have presented a theoretical formalism for evaluating electromagnetic signatures of spacetime fluctuations.
It is based on measuring field correlation tensors at different spacetime locations, and enables to bound the strength 
and the scale of spacetime fluctuations by comparison with experimental data. To that end, we compare the predicted signal to the power and the cross-spectral densities of some of the most sensitive optical Michelson interferometers. 

Crucially, our predicted spectral densities cannot be compared to the experimental data directly, as the predicted signal manifests at zero frequency and where interferometers are not sensitive.
We thus illustrate the principle of constraining models of spacetime fluctuations 
by assuming that the model
spectral density
is present at a frequency at which the interferometer is sensitive. We also provide a correspondence between our bounds and prior single-parameter models of quantum spacetime fluctuations.

Our two-parameter model, capturing the strength and the scale of spacetime fluctuations is able to extract stricter constraints from the  
cross-correlation between the experimental data recorded in the two co-located interferometers, subject to the above caveat. This is evident not only from our two-dimensional exclusion plots in Fig.~\ref{fig:ParamSpace} (b)
but also from the stronger bounds on the $\alpha$ of single-parameter models in Table \ref{tab:comp}. In fact, it is important to note that the bound on $\alpha$ obtained using the CSD measurements is more stringent than that obtained from optical astronomical observations. Subject to the aforementioned caveat, our bound matches that obtained from the shorter wavelength X-ray observations.

The two-dimensional exclusion plots that we introduce should facilitate an unambiguous comparison of various theoretical models of quantum spacetime fluctuations
as well as the experiments being designed to detect them. These could include different laboratory-based 
interferometer configurations~\cite{bentHolo,kwon2022,cardiff21,prad2020} and detection schemes such as counting single photons~\cite{mcculler} 
or detecting higher-order field correlations involving both electric and magnetic fields~\cite{MandelWolf}
as well as  astronomical observations~\cite{NgPerlman2022}.
Finally, our approach is naturally suited to incorporate novel configurations of coupled interferometers using quantum light for quantum gravity tests~\cite{QLtIRB1,QLtIRB2,prad2020}. These could provide new constraints on the structure and nature of spacetime.

Our formalism has already provided insights into experimental routes for detecting quantum spacetime fluctuations. 
These include the role of the separation between two interferometers as well as their arm lengths in constraining the fluctuations based on their CSD. 
The former determines the area of the two-dimensional parameter space that is additionally ruled out by the CSD.
As to the latter, if the arm lengths of the co-located interferometers are not equal to each other, the spacetime fluctuations can manifest as a complex CSD. Complex CSDs at positive frequencies are typical in interferometric experiments which are limited by uncorrelated noise. 

On the theoretical side, future work could identify the specific physical phenomenon that is ignored, due to each assumption or attribute considered when building our model of spacetime fluctuations. 
Further, our approach can be extended to identify the components of metric fluctuations that can accumulate over a distance vis-a-vis those that cannot. We could also examine the role played by the specific autocorrelation function of the spacetime fluctuations~\cite{shar2024}.
Further investigations with other motivated functional dependences of spacetime fluctuations can potentially yield better insight into the frequency dependence and angular correlations of quantum spacetime fluctuations. 
For instance, Appendix \ref{app:gtilde} shows how changing Attribute \ref{mod:metric} can alter the wave equations in our formalism, 
moving beyond the interpretation as a fluctuating refractive index.

\begin{acknowledgments}
We thank H. Grote and K. Dooley for extensive discussions and suggestions crucial to this work, 
as well as Y. Shikano, O. Kwon, R. Roemer, K. Zurek and D. Steeghs for useful discussions. 
BS and AD acknowledge the UK STFC ``Quantum Technologies for Fundamental Physics" program (Grant No. ST/T006404/1) for support. 
SMV acknowledges the support of the Leverhulme Trust under research grant RPG-2019-022.
\end{acknowledgments}

\bibliography{references}

\begin{widetext}

\appendix

\section{\label{app:one} Relativistic wave equation in curved spacetime}

We recall the relativistic wave equation in terms of the electromagnetic field tensor $F_{\alpha \beta}$~\cite{Tsagas2005}
\begin{equation*}
\nonumber \Box F_{\alpha \beta} + 2 R_{\alpha \gamma \beta \delta} F^{\gamma \delta} - R_{\alpha \gamma} F^{\gamma}_{\phantom{\gamma}\beta} + R_{\beta \gamma} F^{\gamma}_{\phantom{\gamma}\alpha} =0,
\label{eqn:relativWEapp}
\end{equation*}
where 
\begin{eqnarray}
\Box F_{\alpha \beta} = g^{\gamma \delta} \nabla_{\gamma} \nabla_{\delta} F_{\alpha \beta} 
=g^{\gamma \delta} \bigg( \partial_{\delta} X_{\alpha \beta \gamma} - \Gamma^{\eta}_{\alpha \delta} X_{\eta \beta \gamma} - \Gamma^{\eta}_{\beta \delta} X_{\alpha \eta \gamma} - \Gamma^{\eta}_{\gamma\delta} X_{\alpha \beta \eta} \bigg),
\end{eqnarray}
with
\be
X_{\alpha \beta \gamma}=\partial_{\gamma} F_{\alpha \beta} - \Gamma^{\eta}_{\beta \gamma} F_{\alpha \eta} - \Gamma^{\eta}_{\alpha \gamma} F_{\eta \beta}, 
\ee
\be
R_{\alpha \beta \gamma \delta} = g_{\alpha \nu} \left[ \partial_{\gamma} \Gamma^{\nu}_{\delta \beta} - \partial_{\delta} \Gamma^{\nu}_{\gamma \beta} + \Gamma^{\nu}_{\gamma \eta} \Gamma^{\eta}_{\delta \beta} - \Gamma^{\nu}_{\delta \eta} \Gamma^{\eta}_{\gamma \beta}\right],
\ee
and
\be
R_{\alpha \beta}= R^{\nu}_{\alpha \nu \beta},
\ee
with the Christoffel symbol 
\be
\Gamma^{\alpha}_{\beta \gamma}=g^{\alpha \mu} (\partial_{\beta} g_{\mu \gamma}+\partial_{\gamma} g_{\beta \mu}-\partial_{\mu} g_{\beta \gamma})/2.
\ee

Using the Attribute \ref{mod:metric} of our model in Sec.~\ref{sec:model} in Eq. \eref{eqn:relativWEapp} and defining $G_{\alpha}=\left( \frac{1}{2 \, n^{2} } \partial_{\alpha} n^{2} \right)$, we find
\begin{subequations}
\begin{eqnarray}
\nonumber g^{\gamma \delta} \partial_{\gamma} \partial_{\delta} F_{0i} - n^{2} \partial_{0} \left[ G_{0} F_{0i} \right] - 2 \, n^{2} \, G_{0} \, \partial_{0} F_{0i} - 2 \, n^{2} \, G_{0}^{2} \, F_{0i} + \sum_{j=1}^{3} \left[ \rule{0cm}{0.75cm} \right. - n^{2} \partial_{0} \left[ \frac{1}{n^{2}} \, G_{j} \, F_{ji} \right] + \partial_{j} \left[ G_{j} \, F_{0i} \right]  - G_{j} \, \partial_{0} F_{ji}  \\ 
- \frac{1}{2 \, n^{2}} \left(\partial_{j}^{2} n^{2} \right)  F_{0i}  - 2 G_{j} \, G_{0} \, F_{ji} + \, \left[ \partial_{j} G_{i} \right] F_{0j} + 4 \,G_{j}^{2}\, F_{0i} \left. \rule{0cm}{0.75cm} \right] = 0, \quad (i=1,2,3) \hspace*{3 em}\\
\nonumber g^{\gamma \delta} \partial_{\gamma} \partial_{\delta} F_{ij} - n^{2} \, \partial_{0} \left[ G_{j} F_{i0} \right] - n^{2} \, \partial_{0} \left[ G_{i} F_{0j} \right] - n^{2} G_{i}\, \partial_{0} F_{0j} - n^{2} G_{j} \, \partial_{0} F_{i0} - 2\, n^{2} \, G_{0} G_{i} \, F_{0j} - 2\, n^{2} \, G_{0} G_{j} \, F_{i0}\\
 -  n^{2} G_{0} \, \partial_{0} F_{ij} - \sum_{q=1}^{3} \left[ \rule{0cm}{0.75cm}\right. \left\{ \rule{0cm}{0.3cm}  \partial_{q} G_{i} \right\}  F_{qj} + \left\{ \rule{0cm}{0.3cm}  \partial_{q} G_{j} \right\}  F_{iq} + G_{q} \, \partial_{q} F_{ij} \left. \rule{0cm}{0.75cm} \right]=0, \quad (i,j=1,2,3) \hspace*{3 em} \label{eqn:magFRWE}
\end{eqnarray}
\label{eqn:FullRWE}
\end{subequations}
with $\alpha,\gamma,\delta=0,1,2,3$. 

Assuming the inequalities in \eref{eqn:fluctIneq}, all terms comprising $G_{\alpha}$, any derivative of $G_{\alpha}$ or $\partial_{j}^{2} n^{2}$ in Eq. \eref{eqn:FullRWE} are neglected. This reduces the wave equation to 
\begin{equation}
g^{\gamma \delta} \partial_{\delta} \partial_{\gamma} F_{\alpha \beta}=0.
\end{equation}

As an aside, for better illustration of the different terms in Eq. \eref{eqn:FullRWE}, we consider the dimensions to reduce to just $(c t,x)$ in place of $(c t,x,y,z)$, with $\partial_{t}$ and $\partial_{x}$ denoting the partial derivatives with respect to $x$ and $t$ respectively. This simplifies Eq. \eref{eqn:FullRWE} to
\begin{eqnarray}
\nonumber -n^{2} \partial_{t}^{2} E_{x} + c^{2} \partial_{x}^{2} E_{x}  - \frac{c^{2}}{2 \, n^{2}} \left(\partial_{x}^{2} n^{2} \right)  E_{x}  - n^{2} \, \left[ \partial_{t}  \left( \frac{1}{2 \, n^{2} } \partial_{t} n^{2} \right)\right] E_{x}    + 2 \, c^{2} \, \left[ \partial_{x} \left( \frac{1}{2 \, n^{2} } \partial_{x} n^{2} \right) \right] E_{x} +  c^{2} \, \left( \frac{1}{2 \, n^{2} } \partial_{x} n^{2} \right) \, \partial_{x} E_{x}  \\ 
- 3 \, n^{2} \, \left( \frac{1}{2 \, n^{2} } \partial_{t} n^{2} \right) \, \partial_{t} E_{x} - 2 \, n^{2} \, \left( \frac{1}{2 \, n^{2} } \partial_{t} n^{2} \right)^{2} \, E_{x}  +4 \, c^{2} \, \left( \frac{1}{2 \, n^{2} } \partial_{x} n^{2} \right)^{2} \, E_{x} = 0. \hspace*{3 em}
\label{eqn:2DRWE}
\end{eqnarray}
Here $E_{x}(x,t)$ is the electric field along the only spatial dimension $x$.
We note that Eq. \eref{eqn:magFRWE} has no equivalent equation when considering only one spatial dimension as there can be no magnetic field in such a universe.
In order to better illustrate the terms involved, we simplify Eq. \eref{eqn:2DRWE} assuming $n(x,t)=1+w(x,t)$ with $w(x,t)\ll 1$ and neglecting $w^{q}$ for $q\geqslant 2$. This reduces Eq. \eref{eqn:2DRWE} to
\begin{eqnarray}
\nonumber \partial_{t}^{2} E_{x} - c^{2} \, \partial_{x}^{2} E_{x} + 2 \, c^{2} \,  w \, \partial_{x}^{2} E_{x}  + 3 \, (1-w) \, \left( \partial_{t} w \right) \, (\partial_{t} E_{x}) - c^{2} (1-3 w)  (\partial_{x} w) (\partial_{x} E_{x})  -  c^{2} \,  (1- 3 w) \left(\partial_{x}^{2} w \right)  E_{x} \\
- c^{2} \,  (1 - 12 w) \left(\partial_{x} w \right)^{2} \, E_{x} +   (1 - w)  \left( \partial_{t}^{2} w \right) \, E_{x} + (1-6 w) \, \left( \partial_{t} w \right)^{2} E_{x} = 0. \hspace*{ 3 em}
\end{eqnarray}
A detailed analysis of the effects of each of these terms involved could potentially reveal interesting signatures of spacetime fluctuation.

\section{\label{app:SDcomput} Computing the fourth-order moment of the output field and power covariance in Holometer}

In this section, we find $\overline{M^{2,2}_{\mathrm{out}} \left(\bm{R};\bm{R}\right)}$ with $\bm{R}=\{(0,y,z,\tau), (\delta_{x},y+\delta_{y},z+\delta_{z},\tau+\delta_{\tau})\}$, where the electric field corresponding to the output of the $p$-interferometer ($p=\textsc{I,II}$) is
\begin{eqnarray}
E^{p}_{\mathrm{out}}(\delta_{x},y+\delta_{y},z+\delta_{z},\tau+\delta_{\tau})= \frac{1}{\sqrt{2}}  \bigg[ E^{(\textsc{c}_{p})}_{y}(\delta_{x},y+\delta_{y},z+\delta_{z},\tau+\delta_{\tau}) - E^{(\textsc{d}_{p})}_{y}(\delta_{x},y+\delta_{y},z+\delta_{z},\tau+\delta_{\tau}) e^{- 2 i \varphi_{\text{off}}} \bigg]. \hspace*{2 em}
\end{eqnarray}
We note that we have invoked Attribute \ref{mod:stat} and $\overline{\mathcal{O}}$ denotes the average of any quantity $\mathcal{O}$ over
realisations of the underlying random process. The fourth-order correlation $M^{2,2}_{\mathrm{out}} \left(\bm{R};\bm{R}\right)$ is written explicitly in terms of $M^{2,2}_{\{\textsc{x,y}\};\{\textsc{x',y'}\}} \left( \bm{R};\bm{R} \right)$ as 
\begin{align}
\nonumber M^{2,2}_{\mathrm{out}} \left(\bm{R};\bm{R}\right)=\frac{1}{4} &\left[ \rule{0cm}{0.7cm} \sum_{\text{X}\in \{\text{C}_{p},\text{D}_{p}\}}\sum_{\text{Y}\in \{\text{C}_{q},\text{D}_{q}\}}M^{2,2}_{\{\textsc{x,y}\};\{\textsc{x,y}\}} \left( \bm{R};\bm{R} \right) - \sum_{\text{X}\in \{\text{C}_{p},\text{D}_{p}\}} M^{2,2}_{\{\textsc{x,c}_{q}\};\{\textsc{x,d}_{q}\}} \left( \bm{R};\bm{R} \right) e^{2 i \varphi_{\text{off}}} \right. \\
\nonumber & - \sum_{\text{X}\in \{\text{C}_{p},\text{D}_{p}\}} M^{2,2}_{\{\textsc{x,d}_{q}\};\{\textsc{x,c}_{q}\}} \left( \bm{R};\bm{R} \right) e^{- 2 i \varphi_{\text{off}}} - \sum_{\text{X}\in \{\text{C}_{q},\text{D}_{q}\}} M^{2,2}_{\{\textsc{c}_{p},\textsc{x}\};\{\textsc{d}_{p},\textsc{x}\}} \left( \bm{R};\bm{R} \right) e^{ 2 i \varphi_{\text{off}}} \\
& \nonumber - \sum_{\text{X}\in \{\text{C}_{q},\text{D}_{q}\}} M^{2,2}_{\{\textsc{d}_{p},\textsc{x}\};\{\textsc{c}_{p},\textsc{x}\}} \left( \bm{R};\bm{R} \right) e^{- 2 i \varphi_{\text{off}}} + M^{2,2}_{\{\textsc{c}_{p},\textsc{c}_{q}\};\{\textsc{d}_{p},\textsc{d}_{q}\}} \left( \bm{R};\bm{R} \right) e^{4 i \varphi_{\text{off}}}\\
& \left.  + M^{2,2}_{\{\textsc{d}_{p},\textsc{d}_{q}\};\{\textsc{c}_{p},\textsc{c}_{q}\}} \left( \bm{R};\bm{R} \right) e^{- 4 i \varphi_{\text{off}}} + M^{2,2}_{\{\textsc{d}_{p},\textsc{c}_{q}\};\{\textsc{c}_{p},\textsc{d}_{q}\}} \left( \bm{R};\bm{R} \right) + M^{2,2}_{\{\textsc{c}_{p},\textsc{d}_{q}\};\{\textsc{d}_{p},\textsc{c}_{q}\}} \left( \bm{R};\bm{R} \right) \rule{0cm}{0.7cm} \right].
\label{eqn:m22DC}
\end{align}
Here $p,q=\text{I, II}$. Further, in the case $p=q$, $\delta_{x}=\delta_{y}=\delta_{z}=0$. We discuss the salient steps involved in computing $M^{2,2}_{\mathrm{out}} \left(\bm{R};\bm{R}\right)$ by listing the steps in computing one of the terms in this moment, such as,
\begin{eqnarray}
\nonumber && \overline{M^{2,2}_{\{\textsc{d}_{p},\textsc{d}_{q}\};\{\textsc{c}_{p},\textsc{c}_{q}\}} \left( \bm{R};\bm{R} \right)}\\
\nonumber &=& \int_{-\infty}^{\infty} \rmd k_{y_1} \int_{-\infty}^{\infty} \rmd k_{z_1} \, \int_{-\infty}^{\infty} \rmd k_{y_2} \int_{-\infty}^{\infty} \rmd k_{z_2} \, \int_{-\infty}^{\infty} \rmd k_{y_3} \int_{-\infty}^{\infty} \rmd k_{z_3} \, \int_{-\infty}^{\infty} \rmd k_{y_4} \int_{-\infty}^{\infty} \rmd k_{z_4} \\
\nonumber &&\overline{\widetilde{E}_{\textsc{s}}^{(\textsc{d}_{p})}(k_{z_1},k_{y_1},0) \widetilde{E}_{\textsc{s}}^{(\textsc{c}_{p}) \ast}(0,k_{y_2},k_{z_2})\widetilde{E}_{\textsc{s}}^{(\textsc{d}_{q})}(k_{z_3},k_{y_3},\delta_{z}) \widetilde{E}_{\textsc{s}}^{(\textsc{c}_{q}) \ast}(\delta_{x},k_{y_4},k_{z_4})} \\ 
\nonumber &&\overline{E_{\textsc{t}}^{(\textsc{d}_{p})}(\tau) E_{\textsc{t}}^{(\textsc{d}_{q})}(\tau+\delta_{\tau}) E_{\textsc{t}}^{(\textsc{c}_{p}) \ast}(\tau) E_{\textsc{t}}^{(\textsc{c}_{q}) \ast}(\tau+\delta_{\tau})} \\
&& e^{-i (k_{y_1} y + k_{z_1} z - k_{y_2} y - k_{z_2} z)} e^{-i (k_{y_3} (y+\delta_{y}) + k_{z_3} (z+\delta_{x}) - k_{y_4} (y+\delta_{y}) - k_{z_4} (z+\delta_{z}))}.
\label{eqn:m22ddcc}
\end{eqnarray}
We note here that we have used the Attribute \ref{mod:indep} in obtaining the above expression.
It is evident from Eq.~\eref{eqn:SolnT} that $\overline{E_{\textsc{t}}^{(\textsc{d}_{p})}(\tau) E_{\textsc{t}}^{(\textsc{d}_{q})}(\tau+\delta_{\tau}) E_{\textsc{t}}^{(\textsc{c}_{p}) \ast}(\tau) E_{\textsc{t}}^{(\textsc{c}_{q}) \ast}(\tau+\delta_{\tau})}=|E_{\textsc{t}}(0)|^{4}$. 

We also find using Eqs. \eref{eqn:SolnS} and \eref{eqn:IFTs},
\begin{eqnarray}
\nonumber &&\overline{\widetilde{E}_{\textsc{s}}^{(\textsc{d}_{p})}(k_{z_1},k_{y_1},0) \widetilde{E}_{\textsc{s}}^{(\textsc{c}_{p}) \ast}(0,k_{y_2},k_{z_2}) \widetilde{E}_{\textsc{s}}^{(\textsc{d}_{q})}(k_{z_3},k_{y_3},\delta_{z}) \widetilde{E}_{\textsc{s}}^{(\textsc{c}_{q}) \ast}(\delta_{x},k_{y_4},k_{z_4})} \\
\nonumber &&=  \frac{1}{4} \,  \overline{ \Phi_{z}(0,0,0) \Phi_{z}(\delta_{x},\delta_{y},\delta_{z}) \Phi_{x}^{\ast}(0,0,0) \Phi_{x}^{\ast}(\delta_{x},\delta_{y},\delta_{z})} \: \widetilde{E}_{\mathrm{in}}(k_{z_1},k_{y_1},0) \,  \widetilde{E}_{\mathrm{in}}^{\ast}(0,k_{y_2},k_{z_2}) \, \widetilde{E}_{\mathrm{in}}(k_{z_3},k_{y_3},0)\\
&&  \widetilde{E}_{\mathrm{in}}^{\ast}(0,k_{y_4},k_{z_4}) \, \exp \left[ \frac{i \Li}{k} \left( k_{y_1}^{2} + k_{z_1}^{2} - k_{y_2}^{2} - k_{z_2}^{2}  \right)\right] \exp \left[ \frac{i \Li}{k}  \left( k_{y_3}^{2} + k_{z_3}^{2}  - k_{y_4}^{2} - k_{z_4}^{2} \right)\right],
\label{eqn:spEmom}
\end{eqnarray}
where 
\begin{subequations}
\begin{eqnarray}
&&\Phi_{z}(\delta_{x},\delta_{y},\delta_{z})=\exp \left[ -i k \int\limits_{0}^{2 \Li} \rmd z' \, \ws(z+\delta_{x},y+\delta_{y},z'+\delta_{z}) \right],\\
&&\Phi_{x}(\delta_{x},\delta_{y},\delta_{z})=\exp \left[ -i k \int\limits_{0}^{2 \Li} \rmd x' \, \ws(x'+\delta_{x},y+\delta_{y},z+\delta_{z}) \right].
\end{eqnarray}
\label{eqn:PhiDefn}
\end{subequations}
Here we have considered the case $\Li_{\textsc{i}}=\Li_{\textsc{ii}}=\Li$, i.e., the co-located interferometers are, in principle, completely identical. This is the reason why the LHS of Eq. \eref{eqn:spEmom} depends on $p,q$ while the RHS shows no such dependence.

Further, $\widetilde{E}_{\mathrm{in}}$ is the Fourier transformed field factor corresponding to the input beam at BS$_{p}$ ($p=\text{I, II}$), while $\widetilde{E}_{\textsc{s}}^{(\textsc{c}_{p})}$ and $\widetilde{E}_{\textsc{s}}^{(\textsc{d}_{p})}$ correspond to the output measured at B$_{p}$. For instance, in the Holometer, we know that $\delta_{i}\approx 0.52$ m and $\Li=40$ m with $\delta_{i}< 0.1 \, \Li$ ($i=x,y,z$). We also note that in the above expressions, arguments such as $z+\delta_{x}$ feature in the definition of $\Phi_{z}$, in the context of field contributions due to the light traversing along $z$-axis in the D$_{p}$ arm. This unusual argument is because of the following reason. We know that the light in D$_{p}$ suffers a reflection at BS$_{p}$ before interference with the light from C$_{p}$ arm. Specifically, the beam cross-section on the $x-y$ plane in D$_{p}$ arm on reflection becomes the $y-z$ plane at the detector, especially with $x\to z$ and $y\to y$ in terms arising due to light propagation in the D$_{p}$ arm.

Denoting the input electric field coming in from port A$_{p}$ by $E_{\text{in}}(x,y,0)$, the electric field $E_{\text{in}}(x,y,0)$ on reflection at BS$_{p}$ into arm C$_{p}$ becomes $E_{\text{in}}(0,y,z)$. We emphasize here that in Eq. \eref{eqn:spEmom}, the factors $\widetilde{E}_{\mathrm{in}}(k_{z_1},k_{y_1},0)$ and $\widetilde{E}_{\mathrm{in}}(k_{z_3},k_{y_3},0)$ are Fourier transforms of $E_{\text{in}}(x,y,0)$ in $x$ and $y$, while $\widetilde{E}_{\mathrm{in}}(0,k_{y_2},k_{z_2})$ and $\widetilde{E}_{\mathrm{in}}(0,k_{y_4},k_{z_4})$ are Fourier transforms of $E_{\text{in}}(0,y,z)$ in $y$ and $z$. 

We then use the Attribute \ref{mod:stat} to simplify $\overline{ \Phi_{z}(0,0,0) \Phi_{z}(\delta_{x},\delta_{y},\delta_{z}) \Phi_{x}^{\ast}(0,0,0) \Phi_{x}^{\ast}(\delta_{x},\delta_{y},\delta_{z})}$ as follows. We first define
\begin{subequations}
\begin{eqnarray}
&&\sigma_{z}(\delta_{x},\delta_{y})= \frac{1}{2 \Li} \int\limits_{0}^{2 \Li} \rmd \zeta_{+} \int\limits_{-\zeta_{+}}^{\zeta_{+}} \rmd \zeta_{-} \, \rho(\delta_{x},\delta_{y},\zeta_{-}),\label{eqn:SigZDefn}\\
&&\sigma_{x}(\delta_{y},\delta_{z})=\frac{1}{2 \Li}\int\limits_{0}^{2 \Li} \rmd \chi_{+} \,\int\limits_{-\chi_{+}}^{\chi_{+}} \rmd \chi_{-} \, \rho(\chi_{-},\delta_{y},\delta_{z}),\label{eqn:SigXDefn}\\
&&\xi_{1}(z) = \int\limits_{0}^{2 \Li} \rmd x' \, \int\limits_{0}^{2 \Li} \rmd z' \, \rho(x'-z,0,z-z'),\label{eqn:Xi1Defn}\\
&&\xi_{2 x}(c_{0},c_{1},c_{2},c_{3})= \int\limits_{0}^{2 \Li} \rmd x \, \int\limits_{0}^{2 \Li} \rmd z \, \rho(x+\delta_{x}-c_{0},c_{2}-c_{1},c_{3}-z),
\label{eqn:Xi2xDefn}\\
&&\xi_{2 z}(c_{0},c_{1},c_{2},c_{3})= \int\limits_{0}^{2 \Li} \rmd x \, \int\limits_{0}^{2 \Li} \rmd z \, \rho(x-c_{0},c_{2}-c_{1},c_{3}-z-\delta_{z}),
\label{eqn:Xi2zDefn}
\end{eqnarray}
\label{eqn:CorrSet}
\end{subequations}
where the $\sigma$'s arise from spatial correlations in the spacetime metric fluctuations within an arm of the interferometer, and the $\xi$'s correspond to spatial correlations of the metric fluctuations between the two arms.

Using Eqs. \eref{eqn:PhiDefn} and \eref{eqn:CorrSet}, we find
\begin{eqnarray}
\nonumber \overline{ \Phi_{z}(0,0,0) \Phi_{z}(\delta_{x},\delta_{y},\delta_{z}) \Phi_{x}^{\ast}(0,0,0) \Phi_{x}^{\ast}(\delta_{x},\delta_{y},\delta_{z})} =\exp \Bigg[ - k^{2} \Gs \bigg\{ \Li \left[2 \sigma_{z}(0,0)  + 2 \sigma_{z}(\delta_{x},\delta_{y}) \right] - 2 \xi_{1}(z) \\
+ \Li \left[ 2 \sigma_{x}(0,0) + 2 \sigma_{x}(\delta_{y},\delta_{z}) \right]  - \xi_{2 z}(z+\delta_{x},y+\delta_{y},y,z)
 - \xi_{2 x}(z,y,y+\delta_{y},z+\delta_{z}) \bigg\} \Bigg].
\label{eqn:PhiMom}
\end{eqnarray}

We recall here from Attribute \ref{mod:gauss} that $\rho(\delta_{x},\delta_{y},\delta_{z})= \exp \left( -\sum\limits_{i=x,y,z} \frac{\delta_{i}^{2}}{2 \ell_{i}^{2}} \right)$.
For instance, $\sigma_{z}$ can be simplified using Attribute \ref{mod:gauss}, to
\begin{eqnarray}
\sigma_{z}(0,0) &&= \left[\sqrt{2 \pi} \,\ell_{z} \, \mathrm{erf} \left( \frac{\sqrt{2} \Li}{\ell_{z}} \right) + \frac{\ell_{z}^{2}}{\Li} \, \left\{\exp \left( - \frac{2 \Li^{2}}{\ell_{z}^{2}} \right)-1 \right\} \right],
\end{eqnarray} 
where $\mathrm{erf}(z)$ is the error function. We recall that the fluctuation scale $N_{i}=2\ell_i/\ell_{\textsc{p}}$ ($i=x,y,z$). We consider an isotropic fluctuation scale $N_{\parallel} \equiv N_{x} = N_{y}= N_{z}=$. Using $N_{\parallel} \ll 2 \Li/\ell_{\textsc{p}}$,
\begin{eqnarray}
\sigma_{z}(0,0)\approx \sqrt{\frac{\pi}{2}} \ell_{\textsc{p}} N_{\parallel}.
\label{eqn:sz}
\end{eqnarray} 
Similar simplifications can be carried out to find 
\begin{eqnarray}
\sigma_{x}(0,0)\approx \sqrt{\frac{\pi}{2}} \ell_{\textsc{p}} N_{\parallel}.
\label{eqn:sx}
\end{eqnarray} 
Further, all $\xi$-type correlations are negligible in comparison to the $\sigma$-type correlations in the limit $N_{\parallel} \ll 2 \Li/\ell_{\textsc{p}}$. This can be seen even from the qualitative argument that the correlation across two different arms should be significantly smaller than the correlations within an arm when the correlation scale $N_{\parallel} \ll 2 \Li/\ell_{\textsc{p}}$.

We also consider the input beam to be a perfect Gaussian beam at BS with the input electric field $E_{\text{in}}(x,y,0)$ coming in from port A$_{p}$ ($p=\text{I, II}$), given by,
\begin{eqnarray}
E_{\text{in}}(x,y,0)=\sqrt{\frac{2}{\pi} \,} \, \frac{z_{R}}{W_{0} \sqrt{z_{R}^{2} + z_{0}^{2}}} \exp \left[ \left( -\frac{i k z_{0} W_{0}^{2} + 2 z_{R}^{2} }{2 W_{0}^{2} (z_{0}^{2}+z_{R}^{2})}  \right) (x^{2} + y^{2}) \right].
\label{eqn:inpGauss}
\end{eqnarray}
Here $W_{0}$ is the beam waist, $z_{0}$ is the position of the beam waist, and $z_{R}=\pi W_{0}^{2}/\lambda$. 

Using Eqs. \eref{eqn:spEmom}, \eref{eqn:PhiMom}, \eref{eqn:sz}, \eref{eqn:sx}, and \eref{eqn:inpGauss} in Eq. \eref{eqn:m22ddcc}, we find 
\begin{eqnarray}
\overline{M^{2,2}_{\{\textsc{d}_{p},\textsc{d}_{q}\};\{\textsc{c}_{p},\textsc{c}_{q}\}} \left( \bm{R};\bm{R} \right)} &=& \, |\mu_{1}(\Li,y,z) \mu_{1}\left(\Li,y+\delta_{s},z+\delta_{s}\right)|^{2} \, \exp \Bigg[ - 2 \sqrt{2 \pi} \mathcal{K} \Li \left\{ 1 + \mathcal{D} \right\} \Bigg].
\end{eqnarray}
Here,
\begin{eqnarray}
&&\mu_{1}(\Li,y,z)  = \sqrt{\frac{2}{\pi}} \frac{k z_{R} W_{0} \sqrt{z_{R}^{2} + z_{0}^{2}}}{(W_{0}^{2} k (z_{R}^{2} + z_{0}^{2}) - 2 i \Li (2 z_{R}^{2} + i k W_{0}^{2} z_{0}) )} \exp \left[ - \frac{k \left( 2 z_{R}^{2} + i k W_{0}^{2} z_{0} \right) \left( y^{2} + z^{2} \right) }{2 \left[ W_{0}^{2} k (z_{R}^{2} + z_{0}^{2}) - 2 i \Li (2 z_{R}^{2} + i k W_{0}^{2} z_{0}) \right]}   \right], \hspace*{0.5 em}
\label{eqn:mu1}\\
&&\mathcal{D}= \frac{1}{2} \left[ \exp \left(-\frac{2 (\delta_{x}^{2}+\delta_{y}^{2})}{ \ell_{\textsc{p}}^{2} N_{\parallel}^{2}}\right) + \exp \left(-\frac{2 (\delta_{y}^{2}+\delta_{z}^{2})}{ \ell_{\textsc{p}}^{2} N_{\parallel}^{2}}\right) \right]. \label{eqn:genDdefn}
\end{eqnarray}
We recall that $\mathcal{K}=k^{2} \Gs \ell_{\textsc{p}} N_{\parallel}$. When $\delta_{x}=\delta_{y}=\delta_{z}=\delta_{s}$, we find $\mathcal{D}=\exp \left(-\frac{4 \delta_{s}^{2}}{ \ell_{\textsc{p}}^{2} N_{\parallel}^{2}}\right)$. In the case $\delta_{x}=\delta_{z}=\delta'_{s}$ and $\delta_{y}=0$, we find $\mathcal{D}=\exp \left(-\frac{2 \delta_{s}^{\prime 2}}{ \ell_{\textsc{p}}^{2} N_{\parallel}^{2}}\right)$. However, we use the definition $\mathcal{D}=\exp \left(-\frac{4 \delta_{s}^{2}}{ \ell_{\textsc{p}}^{2} N_{\parallel}^{2}}\right)$ (recall Eq. \eref{eqn:Ddefn}), and in the case when $\delta_{x}=\delta_{z}=\delta'_{s}$ and $\delta_{y}=0$, we consider $\delta_{s}=\delta'_{s}/\sqrt{2}$. 

Using arguments similar to those used in obtaining $\overline{M^{2,2}_{\{\textsc{d}_{p},\textsc{d}_{q}\};\{\textsc{c}_{p},\textsc{c}_{q}\}} \left( \bm{R};\bm{R} \right)}$, we can compute the other terms in Eq. \eref{eqn:m22DC} to obtain $\overline{M^{2,2}_{\mathrm{out}} \left(\bm{R};\bm{R}\right)} $. Therefore, 
\begin{eqnarray}
\nonumber \overline{M^{2,2}_{\mathrm{out}} \left(\bm{R};\bm{R}\right)} &=&\,  |\mu_{1}(\Li, y,z)|^{2} \, \left|\mu_{1}\left(\Li,y+\delta_{s},z+\delta_{s}\right)\right|^{2} \Bigg( 1 - 2 \exp \left[ - \sqrt{2 \pi} \mathcal{K} \Li \right] \cos 2 \varphi_{\mathrm{off}} \\
&& +  \frac{1}{2} \exp \left[ - 2 \sqrt{2 \pi} \mathcal{K} \Li  \right] \left\{\exp \left[ 2 \sqrt{2 \pi} \mathcal{K} \Li  \mathcal{D} \right]  + \exp \left[ - 2 \sqrt{2 \pi} \mathcal{K} \Li \mathcal{D} \right] \cos 4 \varphi_{\mathrm{off}} \right\} \Bigg).
\label{eqn:2ptCorrHolo}
\end{eqnarray}
The corresponding two-point correlation of power is obtained using
\begin{eqnarray}
\overline{P_{\mathrm{out}}^{i}(\tau) P_{\mathrm{out}}^{j}(\tau+\delta_{\tau})}= \left(\epsilon_{0} c\right)^{2}  \, \int_{A} \mathrm{d}y_{1} \, \mathrm{d}z_{1} \, \int_{A} \mathrm{d}y_{2} \, \mathrm{d}z_{2} \: \overline{M^{2,2}_{\mathrm{out}} \left(\bm{R};\bm{R}\right)},\hspace*{2 em}
\label{eqn:P2ptCorrApp}
\end{eqnarray}
where $\int_{A}$ denotes a surface integral over the beam cross-section. 
We also note that the co-located interferometers in the Holometer or the QUEST experiment are built such that $\delta_{s}\,\gg \sqrt{A}$. For instance, $\delta_{s}= 0.91/\sqrt{3}$ m, $\sqrt{A}= 5 \sqrt{\pi} \times 10^{-3}$ m in the case of the Holometer. 
Therefore, we see that $y_{2}-y_{1}\approx \delta_{s}$ and $z_{2}-z_{1}\approx\delta_{s}$ in the integrals considered above.
Using this in addition to Eqs. \eref{eqn:2ptCorrHolo}, \eref{eqn:P2ptCorrApp}, and \eref{eqn:Covij}, we find
\begin{eqnarray}
\mathrm{Cov}_{i,j}(P_{\mathrm{out}})&=& \frac{P_{0}^{2}}{4} \, \left(\rule{0cm}{0.6cm}\right. \frac{e^{- \sqrt{2 \pi} \mathcal{K} \Li}}{2} \left\{e^{ - \sqrt{2 \pi} \mathcal{K} \Li  \left(1 - 2 \mathcal{D} \right) }\rule{0cm}{0.4cm}\right. + \left. e^{ - \sqrt{2 \pi} \mathcal{K} \Li  \left(1 + 2 \mathcal{D} \right) } \rule{0cm}{0.4cm} \cos 4 \varphi_{\mathrm{off}} \right\}  - e^{- 2 \sqrt{2 \pi} \mathcal{K}  \Li } \cos^{2} 2 \varphi_{\mathrm{off}} \left.\rule{0cm}{0.6cm}\right).
\label{eqn:Cov}
\end{eqnarray}
Using Eq. \eref{eqn:CovXDefn} and carrying out the cosine transform in Eq. \eref{eqn:WienKhin}, we obtain the PSD taking $\mathcal{D}=1$ (as $\delta_{s}=0$) in the above expression. 
The CS (with $\delta_{s}\neq 0$) corresponding to Eq. \eref{eqn:Cov} is obtained easily using Eqs. \eref{eqn:CovXDefn} and \eref{eqn:CSdefn} when $\Li_{\textsc{i}}=\Li_{\textsc{ii}}=\Li$. We note here that when $\Li_{\textsc{i}}=\Li_{\textsc{ii}}=\Li$, $\mathrm{Cov}_{\textsc{i,ii}}(\Delta x)=\mathrm{Cov}_{\textsc{ii,i}}(\Delta x)$ in Eq. \eref{eqn:CSdefn}.

\section{\label{app:LiCorr} PSD and CSD in the limit of long correlation lengths}

Considering $N_{\parallel}=2 \Li/\ell_{\textsc{p}}$ in the case of the Holometer and the QUEST experiment, we find that Eqs. \eref{eqn:SigZDefn}--\eref{eqn:Xi2zDefn} give
\begin{eqnarray}
&&\sigma_{z}(\delta_{s},\delta_{s})=\sigma_{x}(\delta_{s},\delta_{s})=\Li \mathcal{D} \fo,\\
&&\xi_{1}(z) = \Li^{2} \ft,\\
&&\nonumber \xi_{2 x}(z,y,y+\delta_{s},z+\delta_{s}) = \\
&&\hspace*{1 em}\xi_{2 z}(z+\delta_{s},y+\delta_{s},y,z) = \Li^{2} \ft \sqrt{\mathcal{D}},\\
&&\mathrm{with} \: \fo=\sqrt{2 \pi} \, \mathrm{erf}(\sqrt{2}) + e^{-2} -1,\\
&&\mathrm{and} \: \ft=\frac{\pi}{2} \left( \mathrm{erf}(\sqrt{2}) \right)^{2}.
\end{eqnarray}
Using the steps illustrated in Appendix \ref{app:SDcomput}, the PSD in this limit $N_{\parallel}=2 \Li/\ell_{\textsc{p}}$ is
\begin{eqnarray}
S(f)= \mathcal{A} \, &&\left[\rule{0cm}{0.6cm}\right. 1 + \, e^{- 2 (2 \fo -\ft) \ko \Li} \,  \cos 4 \varphi_{\mathrm{off}}  - 2  e^{- (2 \fo -\ft) \ko \Li} \, \cos^{2} 2 \varphi_{\mathrm{off}} \left.\rule{0cm}{0.6cm}\right],
\label{eqn:PSDleqLi}
\end{eqnarray}
and the cross-spectral density is
\begin{eqnarray}
\nonumber CS(f)= \mathcal{A} \, \left[\rule{0cm}{0.6cm}\right. e^{- (2 \fo -\ft) \ko \Li} \, e^{\ko \Li \sqrt{\mathcal{D}} (2 \fo \sqrt{\mathcal{D}} - \ft )}  &&+ \, e^{- (2 \fo -\ft) \ko \Li} \, e^{-\ko \Li \sqrt{\mathcal{D}} (2 \fo \sqrt{\mathcal{D}} - \ft )}  \cos 4 \varphi_{\mathrm{off}} \\
&& - 2  e^{- (2 \fo -\ft) \ko \Li} \, \cos^{2} 2 \varphi_{\mathrm{off}} \left.\rule{0cm}{0.6cm}\right].
\label{eqn:CSDleqLi}
\end{eqnarray}
To first-order in Planck length,
\begin{eqnarray}
S(f) \approx \left(\frac{\lambda}{2 \pi} \right)^{2} \frac{(2 \fo-\ft) \ko \Li}{8 \pi} \delta(f),\\
CS(f) \approx \frac{\sqrt{\mathcal{D}}}{2} (2 \sqrt{D} \fo -\ft) \Gs \Li^{2} \delta(f).
\end{eqnarray}
Comparing $S(f)$ with the measured PSD $S_{\mathrm{meas}}(f)$, we get
\begin{equation}
\Gs \leqslant \frac{\hbar c \lambda}{2 (2 \fo-\ft) \Li^{2} P_{0}}.
\end{equation}
Therefore, $\Gs\leqslant 2 \times 10^{-38}$ in the Holometer, while $\Gs\leqslant 2 \times 10^{-37}$ in the QUEST experiment with 6 dB of squeezing. Comparing $|CS(f)|$ with $C_{\mathrm{max}}$, we find that $\Gs\leqslant 10^{-43}$ in the Holometer, while $\Gs\leqslant 10^{-42}$ in the QUEST experiment. It is clear that the cross-spectral density improves the bound on $\Gs$ for the given $N_{\parallel}$.

\section{\label{app:aLIGOcov} Computing PSD in aLIGO}

In this section, we find $\overline{M^{2,2}_{\text{out}} \left(\bm{R}_{0};\bm{R}_{0}\right)}$ with $\bm{R}_{0}=\{(0,y,z,\tau), (0,y,z,\tau)\}$, where 
\begin{eqnarray}
 &E_{\mathrm{out}}(0,y,z,\tau) = \frac{1}{\sqrt{2}} \sum\limits_{q=1}^{N_{\mathrm{rt}}(\tau)} \sqrt{T_{\textsc{m}} R_{\textsc{m}}^{q-1}} \, \bigg[ E^{(\textsc{c})}_{y}(d_{\textsc{c}},y,z,\tau) - E^{(\textsc{d})}_{y}(d_{\textsc{d}},y,z,\tau) e^{- 2 i \varphi_{\text{off}}} \bigg].
\end{eqnarray}
We note that we have invoked Attribute \ref{mod:stat} and $\overline{\mathcal{O}}$ denotes the average of any quantity $\mathcal{O}$ over
realisations of the underlying random process. As mentioned in Sec. \ref{sec:PSD}, we take $N_{\mathrm{rt}}(\tau)=280$ using the fact that less than 2$\%$ of the input light remains after $280$ round-trips of the light beam in each arm. The fourth-order correlation $M^{2,2}_{\text{out}} \left(\bm{R}_{0};\bm{R}_{0}\right)$ is written explicitly in terms of $M^{2,2}_{\{\textsc{x,y}\};\{\textsc{x',y'}\}} \left( \bm{R}_{\textsc{x,y}};\bm{R}_{\textsc{x',y'}} \right)$ as
\begin{align}
\nonumber M^{2,2}_{\mathrm{out}} \left(\bm{R}_{0};\bm{R}_{0}\right)=&\frac{1}{4} \sum_{\substack{q_{1},q_{2}\\q_{3},q_{4}}=1}^{280} T_{\textsc{m}}^{2} \left(\sqrt{R_{\textsc{m}}}\right)^{q_{1}+q_{2}+q_{3}+q_{4}-4} \left[ \rule{0cm}{0.7cm} \sum_{X\in \{C,D\}}\sum_{Y\in \{C,D\}}M^{2,2}_{\{\textsc{x,y}\};\{\textsc{x,y}\}} \left( \bm{R}_{\textsc{x,y}};\bm{R}_{\textsc{x,y}} \right) \right. \\
\nonumber &- \sum_{X\in \{C,D\}} M^{2,2}_{\{\textsc{x,c}\};\{\textsc{x,d}\}} \left( \bm{R}_{\textsc{x,c}};\bm{R}_{\textsc{x,d}} \right) e^{2 i \varphi_{\text{off}}} - \sum_{X\in \{C,D\}} M^{2,2}_{\{\textsc{x,d}\};\{\textsc{x,c}\}} \left( \bm{R}_{\textsc{x,d}};\bm{R}_{\textsc{x,c}} \right) e^{- 2 i \varphi_{\text{off}}}\\
\nonumber &  - \sum_{X\in \{C,D\}} M^{2,2}_{\{\textsc{c,x}\};\{\textsc{d,x}\}} \left( \bm{R}_{\textsc{c,x}};\bm{R}_{\textsc{d,x}} \right) e^{ 2 i \varphi_{\text{off}}} - \sum_{X\in \{C,D\}} M^{2,2}_{\{\textsc{d,x}\};\{\textsc{c,x}\}} \left( \bm{R}_{\textsc{d,x}};\bm{R}_{\textsc{c,x}} \right) e^{- 2 i \varphi_{\text{off}}}\\
& \nonumber  + M^{2,2}_{\{\textsc{c,c}\};\{\textsc{d,d}\}} \left( \bm{R}_{\textsc{c,c}};\bm{R}_{\textsc{d,d}} \right) e^{4 i \varphi_{\text{off}}} + M^{2,2}_{\{\textsc{d,d}\};\{\textsc{c,c}\}} \left( \bm{R}_{\textsc{d,d}};\bm{R}_{\textsc{c,c}} \right) e^{- 4 i \varphi_{\text{off}}} \\
& \left.  + M^{2,2}_{\{\textsc{d,c}\};\{\textsc{c,d}\}} \left( \bm{R}_{\textsc{d,c}};\bm{R}_{\textsc{c,d}} \right) + M^{2,2}_{\{\textsc{c,d}\};\{\textsc{d,c}\}} \left( \bm{R}_{\textsc{c,d}};\bm{R}_{\textsc{d,c}} \right) \rule{0cm}{0.7cm} \right].
\label{eqn:m22ligo}
\end{align}
Here $\bm{R}_{\textsc{x,y}}=\{(d_{\textsc{x}},y,z,\tau), (d_{\textsc{y}},y,z,\tau)\}$ ($\textsc{x,y}\in \{\textsc{c,d}\}$).
We discuss the salient steps involved in computing $M^{2,2}_{\mathrm{out}} \left(\bm{R};\bm{R}\right)$ by listing the steps in computing one of the terms in this moment, such as,
\begin{eqnarray}
\nonumber && \sum_{\substack{q_{1},q_{2}\\q_{3},q_{4}}=1}^{280} \left(\sqrt{R_{\textsc{m}}}\right)^{q_{1}+q_{2}+q_{3}+q_{4}-4} \: \overline{M^{2,2}_{\{\textsc{d,d}\};\{\textsc{c,c}\}} \left( \bm{R}_{\textsc{d,d}};\bm{R}_{\textsc{c,c}} \right)}\\
\nonumber &=& \int_{-\infty}^{\infty} \rmd k_{y_1} \int_{-\infty}^{\infty} \rmd k_{z_1} \, \int_{-\infty}^{\infty} \rmd k_{y_2} \int_{-\infty}^{\infty} \rmd k_{z_2} \, \int_{-\infty}^{\infty} \rmd k_{y_3} \int_{-\infty}^{\infty} \rmd k_{z_3} \, \int_{-\infty}^{\infty} \rmd k_{y_4} \int_{-\infty}^{\infty} \rmd k_{z_4} \\
\nonumber &&\left(\sum_{\substack{q_{1},q_{2}\\q_{3},q_{4}}=1}^{280} \left(\sqrt{R_{\textsc{m}}}\right)^{q_{1}+q_{2}+q_{3}+q_{4}-4} \: \overline{\widetilde{E}_{\textsc{s}}^{(\textsc{d})}(k_{z_1},k_{y_1},0) \widetilde{E}_{\textsc{s}}^{(\textsc{c}) \ast}(0,k_{y_2},k_{z_2})\widetilde{E}_{\textsc{s}}^{(\textsc{d})}(k_{z_3},k_{y_3},0) \widetilde{E}_{\textsc{s}}^{(\textsc{c}) \ast}(0,k_{y_4},k_{z_4})} \right)\\
&&\overline{\left[E_{\textsc{t}}^{(\textsc{d})}(\tau)\right]^{2} \, \left[E_{\textsc{t}}^{(\textsc{c}) \ast}(\tau)\right]^{2}} e^{-i (k_{y_1} y + k_{z_1} z - k_{y_2} y - k_{z_2} z)} e^{-i (k_{y_3} y + k_{z_3} z - k_{y_4} y - k_{z_4} z)}.
\label{eqn:m22dcligo}
\end{eqnarray}
We note here that we have used the Attribute \ref{mod:indep} in obtaining the above expression.
It is evident from Eq.~\eref{eqn:SolnT} that $\overline{\left[E_{\textsc{t}}^{(\textsc{d})}(\tau)\right]^{2} \, \left[E_{\textsc{t}}^{(\textsc{c}) \ast}(\tau)\right]^{2}}=|E_{\textsc{t}}(0)|^{4}$. 

We also find using Eqs. \eref{eqn:SolnS} and \eref{eqn:IFTs},
\begin{eqnarray}
\nonumber &&\sum_{\substack{q_{1},q_{2}\\q_{3},q_{4}}=1}^{280} \left(\sqrt{R_{\textsc{m}}}\right)^{q_{1}+q_{2}+q_{3}+q_{4}-4} \: \overline{\widetilde{E}_{\textsc{s}}^{(\textsc{d})}(k_{z_1},k_{y_1},0) \widetilde{E}_{\textsc{s}}^{(\textsc{c}) \ast}(0,k_{y_2},k_{z_2}) \widetilde{E}_{\textsc{s}}^{(\textsc{d})}(k_{z_3},k_{y_3},0) \widetilde{E}_{\textsc{s}}^{(\textsc{c}) \ast}(0,k_{y_4},k_{z_4})} \\
\nonumber &&=  \frac{1}{4} \, \sum_{\substack{q_{1},q_{2}\\q_{3},q_{4}}=1}^{280} \left(\sqrt{R_{\textsc{m}}}\right)^{q_{1}+q_{2}+q_{3}+q_{4}-4} \:   \overline{ \Phi_{z}^{(q_{1})}(0,0,0) \Phi_{z}^{(q_{2})}(0,0,0) \left[\Phi_{x}^{(q_{3})}(0,0,0) \Phi_{x}^{(q_{4})}(0,0,0)\right]^{\ast}} \\
\nonumber && \hspace*{10 em} \widetilde{E}_{\mathrm{in}}(k_{z_1},k_{y_1},0) \,  \widetilde{E}_{\mathrm{in}}^{\ast}(0,k_{y_2},k_{z_2}) \, \widetilde{E}_{\mathrm{in}}(k_{z_3},k_{y_3},0) \widetilde{E}_{\mathrm{in}}^{\ast}(0,k_{y_4},k_{z_4}) \, \\
&& \hspace*{10 em} \exp \left[ \frac{i \Li}{k} \left( k_{y_1}^{2} + k_{z_1}^{2} - k_{y_2}^{2} - k_{z_2}^{2}  \right)\right] \,  \exp \left[ \frac{i \Li}{k}  \left( k_{y_3}^{2} + k_{z_3}^{2}  - k_{y_4}^{2} - k_{z_4}^{2} \right)\right],
\label{eqn:spEmomLIGO}
\end{eqnarray}
where 
\begin{subequations}
\begin{eqnarray}
&&\Phi_{z}^{(q)}(0,0,0)=\exp \left[ -i k \int\limits_{0}^{2 q \Li} \rmd z' \, \ws(z,y,z') \right],\\
&&\Phi_{x}^{(q)}(0,0,0)=\exp \left[ -i k \int\limits_{0}^{2 q \Li} \rmd x' \, \ws(x',y,z) \right].
\end{eqnarray}
\label{eqn:PhiDefnLIGO}
\end{subequations}
As pointed out in Appendix \ref{app:SDcomput}, $\widetilde{E}_{\mathrm{in}}$ is the Fourier transformed field factor corresponding to the input beam at BS, while $\widetilde{E}_{\textsc{s}}^{(\textsc{c})}$ and $\widetilde{E}_{\textsc{s}}^{(\textsc{d})}$ correspond to the output measured at B. As before, denoting the input electric field coming in from port A by $E_{\text{in}}(x,y,0)$, the electric field $E_{\text{in}}(x,y,0)$ on reflection at BS into arm C becomes $E_{\text{in}}(0,y,z)$. We emphasize here that in Eq. \eref{eqn:spEmomLIGO}, the factors $\widetilde{E}_{\mathrm{in}}(k_{z_1},k_{y_1},0)$ and $\widetilde{E}_{\mathrm{in}}(k_{z_3},k_{y_3},0)$ are Fourier transforms of $E_{\text{in}}(x,y,0)$ in $x$ and $y$, while $\widetilde{E}_{\mathrm{in}}(0,k_{y_2},k_{z_2})$ and $\widetilde{E}_{\mathrm{in}}(0,k_{y_4},k_{z_4})$ are Fourier transforms of $E_{\text{in}}(0,y,z)$ in $y$ and $z$. 

We then use the Attribute \ref{mod:stat} to simplify $\overline{ \Phi_{z}^{(q_{1})}(0,0,0) \Phi_{z}^{(q_{2})}(0,0,0) \left[\Phi_{x}^{(q_{3})}(0,0,0) \Phi_{x}^{(q_{4})}(0,0,0)\right]^{\ast}}$ as follows. We first define
\begin{subequations}
\begin{eqnarray}
&&\sigma_{z}^{(p,q)}= \frac{1}{2 \Li} \, \int\limits_{0}^{2 p \Li} \rmd z_{1} \int\limits_{0}^{2 q \Li} \rmd z_{2} \, \rho(0,0,z_{2}-z_{1}),\\
&&\sigma_{x}^{(p,q)}= \frac{1}{2 \Li} \, \int\limits_{0}^{2 p \Li} \rmd x_{1} \int\limits_{0}^{2 q \Li} \rmd x_{2} \, \rho(x_{2}-x_{1},0,0),\\
&&\xi_{1}^{(p,q)}(z) = \int\limits_{0}^{2 p \Li} \rmd x' \, \int\limits_{0}^{2 q \Li} \rmd z' \, \rho(x'-z,0,z-z'),
\end{eqnarray}
\label{eqn:CorrSetLIGO}
\end{subequations}
where $p,q \in \mathbb{Z}_{+}$. We also note that the $\sigma$'s arise from spatial correlations in the spacetime metric fluctuations within an arm of the interferometer, and the $\xi$'s correspond to spatial correlations of the metric fluctuations between the two arms.

Using Eqs. \eref{eqn:PhiDefnLIGO} and \eref{eqn:CorrSetLIGO}, we find
\begin{eqnarray}
\nonumber \overline{ \Phi_{z}^{(q_{1})}(0,0,0) \Phi_{z}^{(q_{2})}(0,0,0) \left[\Phi_{x}^{(q_{3})}(0,0,0) \Phi_{x}^{(q_{4})}(0,0,0)\right]^{\ast}} =\exp \Bigg[ - k^{2} \Gs \bigg\{ \Li \left[ \sigma_{z}^{(q_{1},q_{1})}  +  \sigma_{z}^{(q_{2},q_{2})}  + 2 \sigma_{z}^{(q_{1},q_{2})} \right]  \\
- \left( \xi_{1}^{(q_{1},q_{3})}(z) + \xi_{1}^{(q_{1},q_{4})}(z) + \xi_{1}^{(q_{2},q_{3})}(z) + \xi_{1}^{(q_{2},q_{4})}(z)\right) + \Li \left[ \sigma_{x}^{(q_{3},q_{3})} + \sigma_{x}^{(q_{4},q_{4})}+ 2 \sigma_{x}^{(q_{3},q_{4})} \right] \bigg\} \Bigg].
\label{eqn:PhiMomLIGO}
\end{eqnarray}

We recall here from Attribute \ref{mod:gauss} that $\rho(\delta_{x},\delta_{y},\delta_{z})= \exp \left( -\sum\limits_{i=x,y,z} \frac{\delta_{i}^{2}}{2 \ell_{i}^{2}} \right)$. We also use the fluctuation scale $N_{i}=2\ell_i/\ell_{\textsc{p}}$ ($i=x,y,z$) with an isotropic fluctuation scale $N_{\parallel} \equiv N_{x} = N_{y}= N_{z}=$. This helps in simplifying Eq. \eref{eqn:CorrSetLIGO} as follows.
For instance, $\sigma_{z}^{(p,q)}$ can be simplified using Attribute \ref{mod:gauss}, to
\begin{eqnarray}
\nonumber \sigma_{z}^{(p,q)} &&= \sqrt{\frac{\pi}{2}} \,\frac{\ell_{\textsc{p}} N_{\parallel}}{2} \, \left[ p \, \mathrm{erf} \left( \frac{2\sqrt{2} p \Li}{\ell_{\textsc{p}} N_{\parallel}} \right) + q \, \mathrm{erf} \left( \frac{2\sqrt{2} q \Li}{\ell_{\textsc{p}} N_{\parallel}} \right) - (p-q) \, \mathrm{erf} \left( \frac{2\sqrt{2} (p-q) \Li}{\ell_{\textsc{p}} N_{\parallel}} \right) \right]\\
&& - \frac{\ell_{\textsc{p}}^{2} N_{\parallel}^{2}}{8 \Li} \, \left\{1 - \exp \left( - \frac{8 p^{2} \Li^{2}}{N_{\parallel}^{2} \ell_{\textsc{p}}^{2}} \right)- \exp \left( - \frac{8 q^{2} \Li^{2}}{N_{\parallel}^{2} \ell_{\textsc{p}}^{2}} \right)+ \exp \left( - \frac{8 (p-q)^{2} \Li^{2}}{N_{\parallel}^{2} \ell_{\textsc{p}}^{2}} \right) \right\}.
\end{eqnarray} 
Using $\ell_{\textsc{p}} N_{\parallel} / 2 \Li \ll 1$,
\begin{eqnarray}
\sigma_{z}^{(p,q)}\approx \sqrt{\frac{\pi}{2}} \, \text{min}(p,q) \,  \ell_{\textsc{p}} N_{\parallel},
\label{eqn:szLIGO}
\end{eqnarray}
where $\text{min}(p,q)$ chooses the minimum of the two, namely, $p$ and $q$. 
Similar simplifications can be carried out to find 
\begin{eqnarray}
\sigma_{x}^{(p,q)}\approx  \sqrt{\frac{\pi}{2}} \, \text{min}(p,q) \,  \ell_{\textsc{p}} N_{\parallel}.
\label{eqn:sxLIGO}
\end{eqnarray} 
As in Appendix \ref{app:SDcomput}, all $\xi$-type correlations are negligible in comparison to the $\sigma$-type correlations in the limit $N_{\parallel} \ll 2 \Li/\ell_{\textsc{p}}$.

As before, we consider the input beam to be a perfect Gaussian beam at BS with the input electric field $E_{\text{in}}(x,y,0)$ coming in from port A, given by (recalling Eq. \eref{eqn:inpGauss}),
\begin{eqnarray}
E_{\text{in}}(x,y,0)=\sqrt{\frac{2}{\pi} \,} \, \frac{z_{R}}{W_{0} \sqrt{z_{R}^{2} + z_{0}^{2}}} \exp \left[ \left( -\frac{i k z_{0} W_{0}^{2} + 2 z_{R}^{2} }{2 W_{0}^{2} (z_{0}^{2}+z_{R}^{2})}  \right) (x^{2} + y^{2}) \right].
\label{eqn:inpGaussRCL}
\end{eqnarray}
Here we recall that $W_{0}$ is the beam waist, $z_{0}$ is the position of the beam waist, and $z_{R}=\pi W_{0}^{2}/\lambda$. 

Using Eqs. \eref{eqn:spEmomLIGO}, \eref{eqn:PhiMomLIGO}, \eref{eqn:szLIGO}, and \eref{eqn:sxLIGO}, \eref{eqn:inpGaussRCL} in Eq. \eref{eqn:m22dcligo}, we find 
\begin{eqnarray}
\nonumber \sum_{\substack{q_{1},q_{2}\\q_{3},q_{4}}=1}^{280} \left(\sqrt{R_{\textsc{m}}}\right)^{q_{1}+q_{2}+q_{3}+q_{4}-4} \: &&\overline{M^{2,2}_{\{\textsc{d,d}\};\{\textsc{c,c}\}} \left( \bm{R}_{\textsc{d,d}};\bm{R}_{\textsc{c,c}} \right)} = \, \sum_{\substack{q_{1},q_{2}\\q_{3},q_{4}}=1}^{280} \left(\sqrt{R_{\textsc{m}}}\right)^{q_{1}+q_{2}+q_{3}+q_{4}-4} \:  |\mu_{1}(\Li,y,z)|^{4} \, e^{- 4 i \varphi_{\mathrm{off}}}\\
&& \hspace*{7 em} \exp \left[ - \sqrt{\pi / 2} \,  \mathcal{K} \Li \left\{ \sum\limits_{j=1}^{4} q_{j} + 2 \, \text{min}(q_{1},q_{2}) + 2 \, \text{min}(q_{3},q_{4}) \right\} \right].\\
&& \hspace*{7 em} = \, |\mu_{1}(\Li,y,z)|^{4} \, e^{- 4 i \varphi_{\mathrm{off}}} \, \exp\left( -2 \sqrt{2 \pi} \, \mathcal{K} \Li \right) \mathcal{R}_{1}^{2}, \label{eqn:crossterm}
\end{eqnarray}
where
\begin{eqnarray}
\nonumber \mathcal{R}_{1} &=& \left[\rule{0cm}{1.2cm} \right. \frac{1 - \left(\sqrt{R_{\textsc{m}}} e^{-\sqrt{\pi/2} \, \mathcal{K} \Li} \right)^{280}}{\left(1-\sqrt{R_{\textsc{m}}} e^{-\sqrt{\pi/2} \, \mathcal{K} \Li}\right)\left(1-\sqrt{R_{\textsc{m}}} e^{- 3 \sqrt{\pi/2} \, \mathcal{K} \Li}\right)} - \: \frac{\left(\sqrt{R_{\textsc{m}}} e^{-\sqrt{\pi/2} \, \mathcal{K} \Li} \right)^{280} \left[1 - \left(\sqrt{R_{\textsc{m}}} e^{-3 \sqrt{\pi/2} \, \mathcal{K} \Li} \right)^{280}\right]}{\left(1-\sqrt{R_{\textsc{m}}} e^{-\sqrt{\pi/2} \, \mathcal{K} \Li}\right)\left(1-\sqrt{R_{\textsc{m}}} e^{- 3 \sqrt{\pi/2} \, \mathcal{K} \Li}\right)} \\
&& + \sqrt{R_{\textsc{m}}} e^{-\sqrt{\pi / 2} \, \mathcal{K} \Li} \, \frac{\left[1 - \left(R_{\textsc{m}} e^{-2\sqrt{2 \pi} \, \mathcal{K} \Li} \right)^{280}\right]}{\left(1-R_{\textsc{m}} e^{-2\sqrt{2 \pi} \, \mathcal{K} \Li}\right)} \: \left\{\frac{1}{\left(1-\sqrt{R_{\textsc{m}}} e^{- \sqrt{\pi/2} \, \mathcal{K} \Li}\right)} - \frac{e^{-\sqrt{2 \pi} \, \mathcal{K} \Li}}{\left(1-\sqrt{R_{\textsc{m}}} e^{- 3 \sqrt{\pi/2} \, \mathcal{K} \Li}\right)} \right\} \left.\rule{0cm}{1.2cm} \right].
\end{eqnarray}
We also recall that the definition of $\mu_{1}(\Li,y,z)$ is given in Eq. \eref{eqn:mu1} and $\mathcal{K}=k^{2} \Gs \ell_{\textsc{p}} N_{\parallel}$.

Using arguments similar to those used in obtaining Eq. \eref{eqn:crossterm}, we can compute the other terms in Eq. \eref{eqn:m22ligo} to obtain $\overline{M^{2,2}_{\mathrm{out}} \left(\bm{R}_{0};\bm{R}_{0}\right)} $. Defining
\begin{eqnarray}
\nonumber \mathcal{R}_{2} &=& \left[\rule{0cm}{1.2cm} \right. \frac{1 - \left(\sqrt{R_{\textsc{m}}} e^{-\sqrt{\pi/2} \, \mathcal{K} \Li} \right)^{280}}{\left(1-\sqrt{R_{\textsc{m}}} e^{-\sqrt{\pi/2} \, \mathcal{K} \Li}\right)\left(1-\sqrt{R_{\textsc{m}}} e^{\sqrt{\pi/2} \, \mathcal{K} \Li}\right)} - \: \frac{\left(\sqrt{R_{\textsc{m}}} e^{-\sqrt{\pi/2} \, \mathcal{K} \Li} \right)^{280} \left[1 - \left(\sqrt{R_{\textsc{m}}} e^{\sqrt{\pi/2} \, \mathcal{K} \Li} \right)^{280}\right]}{\left(1-\sqrt{R_{\textsc{m}}} e^{-\sqrt{\pi/2} \, \mathcal{K} \Li}\right)\left(1-\sqrt{R_{\textsc{m}}} e^{ \sqrt{\pi/2} \, \mathcal{K} \Li}\right)} \\
&& + \sqrt{R_{\textsc{m}}} \, \frac{\left[1 - \left(R_{\textsc{m}}\right)^{280}\right]}{\left(1-R_{\textsc{m}}\right)} \: \left\{\frac{e^{-\sqrt{\pi / 2} \, \mathcal{K} \Li}}{\left(1-\sqrt{R_{\textsc{m}}} e^{- \sqrt{\pi/2} \, \mathcal{K} \Li}\right)} - \frac{e^{\sqrt{\pi/2} \, \mathcal{K} \Li}}{\left(1-\sqrt{R_{\textsc{m}}} e^{\sqrt{\pi/2} \, \mathcal{K} \Li}\right)} \right\} \left.\rule{0cm}{1.2cm} \right],\\
\mathcal{R}_{3}&=& \frac{1-\left(\sqrt{R_{\textsc{m}}}\right)^{280}}{1-\sqrt{R_{\textsc{m}}}}, \quad \text{and} \\
\mathcal{R}_{4}&=& \frac{1-\left(\sqrt{R_{\textsc{m}}} \, e^{- \sqrt{\pi/2} \, \mathcal{K} \Li}\right)^{280}}{1-\sqrt{R_{\textsc{m}}}\, e^{- \sqrt{\pi/2} \, \mathcal{K} \Li}},
\end{eqnarray}
we find 
\begin{eqnarray}
\nonumber \overline{M^{2,2}_{\mathrm{out}} \left(\bm{R}_{0};\bm{R}_{0}\right)} &=&\,  |\mu_{1}(\Li, y,z)|^{4} \,T_{\textsc{m}}^{2} \, \Bigg( \mathcal{R}_{3}^{4} - 2 \exp \left[ - \sqrt{2 \pi} \mathcal{K} \Li \right] \mathcal{R}_{3}^{2} \mathcal{R}_{4}^{2} \cos 2 \varphi_{\mathrm{off}} \\
&& +  \frac{1}{2} \exp \left[ - 2 \sqrt{2 \pi} \mathcal{K} \Li  \right] \left\{\exp \left[ 2 \sqrt{2 \pi} \mathcal{K} \Li  \right] \mathcal{R}_{2}^{2}  + \exp \left[ - 2 \sqrt{2 \pi} \mathcal{K} \Li \right] \mathcal{R}_{1}^{2} \cos 4 \varphi_{\mathrm{off}} \right\} \Bigg).
\label{eqn:2ptCorrLIGO}
\end{eqnarray}
Using Eqs. \eref{eqn:2ptCorrLIGO}, \eref{eqn:P2ptCorr}, and \eref{eqn:Covij}, we find
\begin{eqnarray}
\mathrm{Cov}_{i,j}(P_{\mathrm{out}})&=& \frac{P_{0}^{2}}{4} \,T_{\textsc{m}}^{2} \, \left(\rule{0cm}{0.6cm}\right. \frac{e^{- 2 \sqrt{2 \pi} \mathcal{K} \Li}}{2} \left\{e^{ 2 \sqrt{2 \pi} \mathcal{K} \Li } \, \mathcal{R}_{2}^{2} \rule{0cm}{0.4cm}\right. + \left. e^{ - 2 \sqrt{2 \pi} \mathcal{K} \Li} \, \mathcal{R}_{1}^{2}  \rule{0cm}{0.4cm} \cos 4 \varphi_{\mathrm{off}} \right\}  - e^{- 2 \sqrt{2 \pi} \mathcal{K}  \Li } \mathcal{R}_{4}^{4} \cos^{2} 2 \varphi_{\mathrm{off}} \left.\rule{0cm}{0.6cm}\right).
\label{eqn:CovLIGO}
\end{eqnarray}
Using Eq. \eref{eqn:CovXDefn} and carrying out the cosine transform, we obtain the PSD from the above expression as,
\be
S(f)=\mathcal{A} T_{\textsc{m}}^{2} \left[ \mathcal{R}_{2}^{2} + \, \mathcal{B}^2 \mathcal{R}_{1}^{2}  \cos 4 \varphi_{\mathrm{off}} - 2 \mathcal{B} \mathcal{R}_{4}^{4} \cos^{2} 2 \varphi_{\mathrm{off}} \right]
\ee
Here,
\begin{eqnarray}
&&\mathcal{A}= \frac{1}{4 \pi}\left(\frac{\lambda}{8 \pi \varphi_{\mathrm{off}}}\right)^{2} \, \Theta(f),\\
&& \mathcal{B} = \exp{\left(- 2 \sqrt{2 \pi} \mathcal{K} \Li \right)},\\
\end{eqnarray}
To first-order in Planck length, i.e., $\mathcal{B}=1- 2 \sqrt{2 \pi} \mathcal{K} \Li$, the PSD becomes
\begin{equation}
S(f) \approx \frac{\mathcal{K} \Li}{2 \sqrt{2 \pi}} \left(\frac{\lambda T_{\textsc{m}}}{2 \pi}\right)^{2}  \,  \left(\frac{1- (\sqrt{R_{\textsc{m}}})^{280}}{1- \sqrt{R_{\textsc{m}}}}\right)^{4} \Theta(f).
\label{eqn:psdLIGOapp}
\end{equation}
Here we have used the fact that $\varphi_{\mathrm{off}}\ll 1$, as in the setups considered $\varphi_{\mathrm{off}}\approx 10^{-5}$ rad (see Table \ref{tab:param}). 
To first-order in Planck length, we have taken $\mathcal{R}_{1}$, $\mathcal{R}_{2}$, and $\mathcal{R}_{4}^{2}$ to be approximately equal to $\mathcal{R}_{3}^{2}$.

\section{\label{app:gtilde} Effects of the chosen spacetime metric}

Attribute \ref{mod:metric} is an important choice made on how the spacetime metric fluctuates. If we choose another metric, for instance,
\begin{eqnarray}
&&\nonumber \widetilde{g}^{00}=-1, \quad \widetilde{g}^{11}=n^{2}(x,y,z,t), \\
&&\widetilde{g}^{22}=\widetilde{g}^{33}=1,
\end{eqnarray}
with all off-diagonal elements set to zero. Here $n(x,y,z,t)$ is a random field in the 4-dimensional spacetime with $n(x,y,z,t)=n_{0}(y,z,t) n_{1}(x)$. The inequalities listed in \eref{eqn:fluctIneq} are used in what follows. The Assumptions \ref{Ass:VarSep}--\ref{Ass:Parax} are suitably modified according to the chosen form of $n(x,y,z,t)$ and $E_{i}(x,y,z,t)$ in a straightforward manner. Now the SDEs are listed below for three different possible directions of light propagation.

\noindent \textit{Case X:} If light propagates along $x$-axis and if the electric field is of the form $E_{i}(x,y,z,t)=E_{1}(x) E_{0}(y,z,t) e^{-i (k x - \Omega t)}$, the SDEs are
\begin{eqnarray}
\frac{1}{c} \, \frac{\partial E_{0}}{\partial t} &=& -\frac{i}{2 k} \left( \frac{\partial^{2} E_{0}}{\partial y^{2}} + \frac{\partial^{2} E_{0}}{\partial z^{2}} \right) + \frac{i k}{2} (n_{0}^{2}-1) E_{0}, \:\:\\
\frac{\partial E_{1}}{\partial x} &=& \frac{i k}{2} E_{1} -\frac{i k}{2 n_{1}^{2}} E_{1}.
\end{eqnarray}
In this case, we clearly see that there is a phase accumulated as light propagates in time $t$. However, due to the similarity in the structure of the SDEs to those discussed in Sec. \ref{sec:EMsig}, we can see that the phase difference becomes independent of metric fluctuations along $x$.

\noindent \textit{Case Y:} If light propagates along $y$-axis and if the electric field is of the form $E_{i}(x,y,z,t)=E_{1}(x) E_{0}(y,z,t) e^{-i (k y - \Omega t)}$, the SDEs are
\begin{eqnarray}
\frac{\partial E_{0}}{\partial y}+ \frac{1}{c} \, \frac{\partial E_{0}}{\partial t} &=& -\frac{i}{2 k} \frac{\partial^{2} E_{0}}{\partial z^{2}} + \frac{i k}{2} n_{0}^{2} E_{0}, \label{eqn:SDEY1} \:\:\\
\frac{\partial^{2} E_{1}}{\partial x^{2}} &=& -\frac{k^{2}}{n_{1}^{2}} E_{1}.\label{eqn:SDEY2}
\end{eqnarray}
Equation \eref{eqn:SDEY1} points to a phase fluctuation along $y$ and in time $t$. In addition to this, $k^{2}/n_{1}^{2}$ in Eq. \eref{eqn:SDEY2} suggests a fluctuating wavelength. This agrees with the earlier work (see, for instance,~\cite{NgPerlman2022}) that point to both fluctuations in phase difference between the arms of the interferometer and fluctuations in the wavelength of the light. However, the SDEs corresponding to this case are not readily solvable analytically, in contrast to those considered in Sec. \ref{sec:EMsig}. So, in this preliminary investigation, we confine our computations to the metric assumed in Attribute \ref{mod:metric}.

\noindent \textit{Case Z:} If light propagates along $z$-axis, the results are identical to that obtained in Case Y.

As expected, the choice of the metric is crucial on how the spacetime fluctuations affect the observables of interest. So, this investigation reveals the need to further examine the precise effects of the different metric fluctuations in detail.

\end{widetext}

\end{document}